\documentclass[singlecolumn]{article}

\usepackage{arxiv}

\usepackage[utf8]{inputenc} 
\usepackage[T1]{fontenc}    
\usepackage{hyperref}       
\usepackage{url}            
\usepackage{booktabs}       
\usepackage{amsfonts}       
\usepackage{nicefrac}       
\usepackage{microtype}      
\usepackage{graphicx}
\usepackage[sort]{natbib}
\setcitestyle{comma,numbers,open={[},close={]}}
\usepackage{doi}

\usepackage{stmaryrd}
\usepackage{pifont} 
\newcommand{\cmark}{\ding{51}}
\newcommand{\xmark}{\ding{55}}
\usepackage{siunitx}
\usepackage[scr=dutchcal,scrscaled=1.05]{mathalfa} 
\usepackage{upgreek}
\usepackage{subfig}

\title{A data-driven multiscale scheme for anisotropic finite strain magneto-elasticity}

\author{
	Heinrich T. Roth\\
	Chair of Computational and\\
	Experimental Solid Mechanics\\
	TU Dresden,
	01062 Dresden, Germany \\
	\And
	Philipp Gebhart\\
	Chair of Mechanics of\\
	Multifunctional Structures\\
	TU Dresden,
	01062 Dresden, Germany \\
	\And
	Karl A. Kalina\\
	Chair of Computational and\\
	Experimental Solid Mechanics\\
	TU Dresden,
	01062 Dresden, Germany \\
	\And
	Thomas Wallmersperger\\
	Chair of Mechanics of\\
	Multifunctional Structures\\
	TU Dresden,
	01062 Dresden, Germany \\
	\And
	Markus K\"{a}stner\thanks{Corresponding author, email: \texttt{markus.kaestner@tu-dresden.de}.} \\
	Chair of Computational and\\
	Experimental Solid Mechanics\\
	TU Dresden, 
	01062 Dresden, Germany \\
}

\renewcommand{\shorttitle}{A data-driven multiscale scheme for anisotropic finite strain magneto-elasticity}

\hypersetup{
pdftitle={Preprint_RothEtAl_2025},
pdfsubject={A data-driven multiscale scheme for anisotropic finite strain magneto-elasticity},
pdfauthor={Roth},
pdfkeywords={Finite strain magneto-elasticity, Magnetorheological elastomers, Physics-augmented neural networks, Constitutive modeling, Anisotropy, Computational homogenization, Finite Element Method},
}

\usepackage{amssymb}
\usepackage{amsthm}
\usepackage{amsmath}
\usepackage{siunitx}
\usepackage{mathtools}		
\mathtoolsset{centercolon}	
\usepackage{newtxmath}      
\usepackage{lineno}
\usepackage{pifont}
\usepackage[dvipsnames]{xcolor}
\usepackage{bbm}
\usepackage{multirow}

\newcommand{\ve}[1]{\boldsymbol{#1}} 
\newcommand{\te}[1]{\boldsymbol{\mathrm{#1}}} 
\newcommand{\Ve}[1]{\underline{\boldsymbol{#1}}} 
\newcommand{\pd}[2]{\frac{\partial #1}{\partial #2}} 
\newcommand{\jump}[1]{\llbracket #1\rrbracket}
\newcommand{\periodplus}[1]{\llparenthesis #1\rrparenthesis}
\newcommand{\dd}{\mathrm{d}}

\newcommand{\op}[1]{\operatorname{#1}}

\newcommand{\norm}[1]{\lvert #1 \rvert}
\newcommand{\frob}[1]{\left\lVert #1 \right\rVert}
\newcommand{\ho}[1]{\bigl< #1 \bigr>}
\newcommand{\samp}[1]{#1^\text{samp}}

\newcommand{\grad}{\nabla}
\newcommand{\Grad}{\nabla_{\!\!X}}
\renewcommand{\div}{\nabla\cdot}
\newcommand{\Div}{\nabla_{\!\!X}\cdot}
\newcommand{\rot}{\nabla\times}
\newcommand{\Rot}{\nabla_{\!\!X}\times}

\newcommand{\Ptot}{\te{P}^\text{tot}}
\newcommand{\sigtot}{\te{\upsigma}^\text{tot}}
\newcommand{\sig}{\upsigma}

\definecolor{myGreen}{rgb}{0,0.5,0}
\definecolor{myYellow}{rgb}{1,0.8,0}

\theoremstyle{definition} 
\newtheorem{remark}{Remark}
\usepackage[font=small]{caption}

\begin{document}

\maketitle

\begin{abstract}
In this work, we develop a neural network-based, data-driven, decoupled multiscale scheme for the modeling of structured magnetically soft magnetorheological elastomers (MREs). 
On the microscale, sampled magneto-mechanical loading paths are imposed on a representative volume element containing spherical particles and an elastomer matrix, and the resulting boundary value problem is solved using a mixed finite element formulation.
The computed microscale responses are homogenized to construct a database for the training and testing of a macroscopic physics-augmented neural network model.
The proposed model automatically detects the material’s preferred direction during training and enforces key physical principles, including objectivity, material symmetry, thermodynamic consistency, and the normalization of free energy, stress, and magnetization.
Within the range of the training data, the model enables accurate predictions of magnetization, mechanical stress, and total stress.
For larger magnetic fields, the model yields plausible results.
Finally, we apply the model to investigate the magnetostrictive behavior of a macroscopic spherical MRE sample, which exhibits contraction along the magnetic field direction when aligned with the material’s preferred direction.
\end{abstract}



\section{Introduction}\label{Introduction}

Due to their unique properties and the challenges associated with their modeling and simulation, magnetoactive materials remain the subject of ongoing research.
One prominent class of such materials is magnetorheological elastomers (MREs), which are composite materials consisting of a polymer matrix and embedded magnetizable particles \cite{Rigbi83,Bastola20,Cantera17,Pelteret2020}.
Due to the significantly lower magnetizability of the matrix compared to the particles, the magnetic behavior of an MRE is governed solely by the shape, distribution and constitutive behavior of the particles.
The micron-sized particles can be magnetically soft, i.e. non-dissipative and fully reversible, as for carbonyl iron \cite{Schümann23,Danas12,Guan07}, or magnetically hard as for NdFeB \cite{Schliephake23,Schümann23,Moreno-Mateos22}, which leads to energy dissipation and path-dependence \cite{Jiles91}.
The corresponding composite materials are referred to as soft-magnetic MREs (sMREs) and hard-magnetic MREs (hMREs), respectively.
MREs are manufactured by adding the magnetizable filler to the non-crosslinked matrix material.
Typical filler particles have diameters in the range of \SIrange{3}{7}{\micro\metre} for sMREs \cite{Bastola20} and \SIrange{5}{100}{\micro\metre} for hMREs \cite{Schliephake23}.
After a chemical process called \textit{curing}, the polymer matrix becomes solid.
Commonly used polymers include silicone rubber and natural rubber.
Since both the matrix and particles exhibit isotropic mechanical and magnetic properties, an MRE with homogeneously distributed spherical particles also shows an isotropic effective behavior \cite{Pelteret2020}.
However, if a magnetic field is applied during the curing process, the particles arrange in chains \cite{Jolly96,Danas12,Hiptmair15}, which causes an anisotropic macroscopic behavior.
Such an MRE will be referred to as a \textit{structured MRE} \cite{Günther12} in the following.
The direction in which the particle chains are aligned is called the \emph{preferred direction} of the material.

The most relevant magneto-mechanical coupling effects of MREs are (i) the magnetostrictive effect, which leads to magnetically induced deformation \cite{Martin06}, and (ii) the magnetorheological effect, which causes magnetically induced stiffness changes \cite{Varga06}.
Both effects are more pronounced for a more compliant matrix \cite{Tasin23,Awietjan12} and for structured MREs \cite{Guan07,Schubert15}.

Applications of the magnetostrictive effect include actuators \cite{Schubert14Diss,Becker2019}, sensors \cite{Du12,Tian11,Volkova17}, and valves \cite{Boese12}. The magnetorheological effect can be utilized in tunable vibration absorbers and dampers \cite{Ginder01,lin06,Kim06,Li13}.
For further information, the reader is referred to, e.g., \cite{Bastola20}.

\subsection{Modeling approaches for MREs}\label{sec:Modeling_approaches_for_MREs}

Models describing the behavior of MREs are categorized as either microscopic or macroscopic, depending on the scale at which they represent the material \cite{Metsch16,Cantera17,Kalina24}.

\textit{Microscopic models} explicitly resolve the heterogeneous microstructure. One class of such models are particle interaction models \cite{Cremer15,Romeis16,Ivaneyko12}, which are efficient, but limited to low particle volume fractions.
Another approach is to explicitly resolve the mechanical and magnetic fields using a continuum formulation \cite{Castaneda11,Kalina17,Danas17}, which allows for the accurate description of materials with arbitrary particle shapes, sizes and volume fractions.
However, solving the boundary value problem (BVP) of a real-world sample with such a model using the Finite Element Method (FEM) would require enormous computing power.
In order to reduce computational costs, homogenization schemes \cite{Chatzigeorgiou14} are employed, in which only a small part of the entire microstructure, a so-called representative volume element (RVE), is used for the macroscopic calculations.
In a coupled multiscale scheme ($\text{FE}^2$) \cite{Keip16,Zabihyan20}, a microscopic RVE problem is solved at each macroscopic integration point without the need to explicitly formulate a macroscopic constitutive model.

\textit{Macroscopic models}, on the other hand, do not explicitly resolve heterogeneous microstructures, but consider a homogeneous material with an equivalent effective macroscopic material behavior.
Phenomenological models purely based on experimental data are available for isotropic magneto-elasticity \cite{Bustamante11, Dorfmann04}, transversely isotropic magneto-elasticity \cite{Bustamante10, Danas12, Itskov16}, and magneto-viscoelasticity \cite{Saxena13, Haldar16}.
These trained macroscopic models are often influenced by the model geometry \cite{Keip17}.
An approach to creating a macroscopic model that is independent of any geometry influences is based on the above-mentioned concept of homogenization.
This means, that an RVE is used to generate a dataset, which is then used to train a macroscopic material model.
After the training, the macroscale model can be used in Finite Element (FE) simulations as a surrogate for the RVE's homogenized response without the need to solve the microscale BVP, which is called decoupled multiscale scheme \cite{Kalina20-2,Gebhart22_2}.
Naturally, this method is much more efficient than the coupled multiscale scheme.
However, one drawback is the difficulty in formulating an accurate macroscopic model.

Due to their high degree of flexibility, \emph{neural networks (NNs)} are a promising alternative to classical approaches for constructing macroscopic models.
Early models of this type \cite{Ghaboussi90,Furukawa99,Theocaris95} lacked physical constraints.
This leads to a large amount of required training data and a poor extrapolation behavior \cite{Rosenkranz2023}.
For this reason, so-called \textit{physics-informed} \cite{Raissi19,Henkes22}, \textit{physics-augmented} \cite{Klein22,Linden23,Rosenkranz2023} or \textit{thermodynamics-based} \cite{Masi21} neural networks have gained in popularity.
Physical constraints can be incorporated in a strong sense by adapting the model's architecture, or in a weak sense by incorporating additional terms to the loss function.
Many models address purely mechanical problems, such as hyperelasticity \cite{Liu2020,Kalina22,Linden23}, viscoelasticity \cite{Rosenkranz2023} and elasto-plasticity \cite{Fuhg2023}.
For coupled problems, however, only models for thermo-elasticity \cite{Zlatic23,Fuhg24}, electro-elasticity \cite{Klein22,Klein24} and magneto-elasticity \cite{Kalina24} are known to the authors.
The model by Kalina~et~al.~\cite{Kalina24}, which is based on physics-augmented neural networks (PANNs), is able to very accurately predict homogenization data, but restricted to isotropic materials and only trained and tested using plane strain and two-dimensional magnetic induction states.

\subsection{Contributions of this work}\label{sec:Contributions_of_this_work}

In this work, we extend the model of Kalina~et~al.~\cite{Kalina24} to transverse isotropy, train it with homogenization data from general three-dimensional load states, and apply it in macroscopic simulations.
To this end, a decoupled data-driven multiscale scheme is employed:
Starting with the microscale material laws, microscopic fields from RVE simulations are homogenized to create a macroscopic dataset.
This dataset serves as the basis for training a macroscale constitutive model, which is then used in macroscopic FE simulations.
By using PANNs, several physical constraints are satisfied per construction, including thermodynamic consistency, material objectivity and symmetry, balance of angular momentum, the volumetric growth condition and zero stress and magnetization in the unloaded state.
Additional conditions on the magnetization curve are imposed in a weak form during the training.
To demonstrate the feasibility of the framework, we generate an approximately transversely isotropic data set via computational homogenization.
After training, the mechanical and magnetic anisotropy and magnetic extrapolation capabilities of the PANN model are investigated.
This is followed by a macroscale simulation to determine the magnetostriction of a spherical MRE sample, the constitutive behavior of which is defined by the PANN model. 
For validation, we compare predictions from the macroscopic model with homogenization results from RVE microscale simulations.
Assumptions made include a purely elastic mechanical behavior and magnetically soft particles.
Furthermore, electric fields and current densities are neglected, and all processes are considered quasi-stationary.

The paper is outlined as follows: In Section~\ref{chap:magneto-hyperelasticity}, the fundamental equations of magneto-mechanics, a homogenization scheme and a variational formulation of the problem are given.
Section~\ref{chap:Data-driven_multiscale_scheme} presents a multiscale scheme that spans data generation, the formulation and training of a macroscale PANN model, and its subsequent application in macroscale simulations.
The proposed approach is exemplarily applied to a specific microstructure in Section~\ref{chap:Numerical_examples}.
We conclude this work with a summary of the results and a discussion of future research directions in Section~\ref{chap:Conclusion}.

\subsection{Notation}\label{sec:Notation}
In the following, scalars are given by italic letters, i.e., $a \in \mathbb{R}$, whereas tensors of order one and two are denoted by boldface italic letters and boldface upright letters, respectively, i.e., $\ve{a} = a_i \ve{e}_i \in \mathcal{L}_1$ and $\te{A} = A_{ij} \ve{e}_i \otimes \ve{e}_j \in \mathcal{L}_2$, with the Cartesian basis vectors $\ve{e}_i \in \mathcal{L}_1$ and the dyadic product $\otimes$. The space of tensors of order $n \in \mathbb{N}_{>1}$ is denoted as $\mathcal{L}_n$.
Unless stated otherwise, the Einstein summation convention is used.
Single contraction, i.e., $\ve{a} \cdot \ve{b} = a_i b_i$ and $\te{A} \cdot \te{B} = A_{ij} B_{jk} \ve{e}_i \otimes \ve{e}_k$, double contraction, i.e., $\te{A} : \te{B} = A_{ij} B_{ji}$, and the cross product, i.e., $\ve{a} \times \ve{b} = e_{ijk} a_i b_j \ve{e}_k$, are denoted as shown, where the Levi-Civita symbol $e_{ijk}$ contains the coordinates of the permutation pseudo tensor $\mathfrak{e} = e_{ijk} \ve{e}_i \otimes \ve{e}_j \otimes \ve{e}_k \in \mathcal{L}_3$.
The trace, determinant, and cofactor are denoted as $\op{tr}(\te{A}) = A_{ii}$, $\op{det}(\te{A})$ and $\op{cof}(\te{A}) = \op{det}(\te{A}) \te{A}^\text{-T}$, respectively.
Within the Euclidean vector space $\mathbb{R}^3$, subsets of $\mathcal{L}_2$ include the set of symmetric second order tensors $\mathscr{Sym} = \{\te{A} \in \mathcal{L}_2 \ |\ \te{A} = \te{A}^\text{T}\}$, 
the set of unimodular tensors $\mathscr{Uni} = \{\te{A} \in \mathcal{L}_2 \ |\ \op{det}\te{A} = 1 \}$, the orthogonal group $\mathcal{O}(3) = \{ \te{A} \in \mathcal{L}_2 \ |\ \te{A} \cdot \te{A}^\text{T} = \te{1}\}$, the special orthogonal group $\mathcal{SO}(3) = \{ \te{A} \in \mathcal{L}_2 \ |\ \te{A} \cdot \te{A}^\text{T} = \te{1}, \op{det}(\te{A}) = 1\}$ and the set of invertible second order tensors with a positive determinant $\mathcal{GL}^+(3) = \{\te{A} \in \mathcal{L}_2 \ |\ \op{det}(\te{A}) > 0\}$.
Here $\te{1} = \delta_{ij} \ve{e}_i \otimes \ve{e}_j$ is the identity tensor of order two with the Kronecker delta $\delta_{ij}$.
The set of unit vectors is given as $\mathcal{N} = \{\ve{n} \in \mathcal{L}_1 \ |\ \norm{\ve{n}} = 1\}$ with the Euclidean norm $\norm{\ve{a}} = \sqrt{a_i a_i}$.
The Frobenius norm is defined as $\frob{\te{A}} = \sqrt{A_{ij} A_{ij}}$.
Spatial derivatives of tensors of order 1 or 2 can be defined using the nabla operator $\nabla$ as $\nabla_X \circ \te{A} = \ve{e}_i \circ \pd{\te{A}}{X_i}$ with the coordinates in the reference configuration $X_i$ and  $\nabla \circ \te{A} = \ve{e}_i \circ \pd{\te{A}}{x_i}$ with the coordinates in the current configuration $x_i$.
Depending on the symbol chosen for the placeholder $\circ \in \{ \ , \cdot, \times \}$, the expression is called gradient, divergence, or curl, respectively.
The jump operator $\jump{\circ} = (\circ)^+ - (\circ)^-$ is used to denote the jump of a variable $\circ$ across material surfaces $\mathcal{S}_0$ or $\mathcal{S}$ from subdomains $\mathcal{B}^-_0$ to $\mathcal{B}^+_0$ or $\mathcal{B}^-$ to $\mathcal{B}^+$.
To denote the differences of a variable on two opposite boundaries $\partial\mathcal{B}^{\text{RVE,}-}_0$ and $\partial\mathcal{B}^{\text{RVE,}+}_0$ of an RVE, the jump operator $\periodplus{(\circ)} = (\circ)^+ - (\circ)^-$ is defined.
For readability, function arguments may be omitted throughout this work, and the same symbol is used for both a function and its value.

\section{Finite strain magneto-hyperelasticity}\label{chap:magneto-hyperelasticity}

\subsection{Kinematics}\label{chap:Kinematics}

At the time $t_0 \in \mathbb{R}_{\ge 0}$, a deformable body is given in its reference configuration $\mathcal{B}_0 \subset \mathbb{R}^3$, where no loads act on it.
The body is in an undeformed state and can be described with the referential coordinates $\ve{X} \in \mathcal{L}_1$.
At the time $t \in \mathcal{T}$, the body is in its current configuration $\mathcal{B} \subset \mathbb{R}^3$ with the current coordinates $\ve{x} \in \mathcal{L}_1$ and $\mathcal{T} = \{ t \in \mathbb{R} \mid t \ge t_0 \}$. 
The motion of the body is described with the deformation map $\ve{\varphi} : \mathcal{B}_0 \times \mathcal{T} \to \mathcal{B}, (\ve{X},t) \mapsto \ve{x}$.
The difference between current and reference coordinates is the displacement vector $\ve{u}(\ve{X},t) = \ve{\varphi}(\ve{X},t) - \ve{X} \in \mathcal{L}_1$.
The velocity $\ve{v} = \dot{\ve{u}} \in \mathcal{L}_1$ is defined using the material time derivative $(\dot{\circ})$.
By differentiation, the deformation gradient $\te{F} = (\Grad \ve{\varphi})^\text{T} = \te{1} + (\Grad \ve{u})^\text{T} \in \mathcal{GL}^+(3)$ and its determinant, the Jacobian determinant $J = \op{det}(\te{F}) \in \mathbb{R}_{> 0}$ are obtained.
The polar decomposition of $\te{F} = \te{R} \cdot \te{U}$ yields the special orthogonal rotation tensor $\te{R} \in \mathcal{SO}(3)$ and the positive definite symmetric right stretch tensor $\te{U} \in \mathscr{Sym}$, $\ve{x} \cdot \te{U} \cdot \ve{x} > 0 \ \forall \ve{x} > \ve{0} \in \mathcal{L}_1 \backslash \{\ve{0}\} $ with the eigenvalues of $\te{U}$ being denoted as principal stretches $\lambda_i \in \mathbb{R}_{>0}, i \in \{1,2,3\}$.
Further symmetric deformation measures include the right Cauchy-Green deformation tensor $\te{C} = \te{U}^2 = \te{F}^\text{T} \cdot \te{F} \in  \mathscr{Sym} \cap \mathcal{GL}^+(3) $ and the Green-Lagrange strain tensor $\te{E} = \frac{1}{2} (\te{C} - \te{1}) \in \mathscr{Sym} \cap \mathcal{GL}^+(3) $ \cite{Haupt2002,Holzapfel2002,Altenbach}.

\subsection{Maxwell and balance equations}\label{sec:Maxwell_and_balance_equations}

In this work, we assume quasistatic processes and the absence of electric fields and currents.
Under these conditions, the Maxwell equations in the current configuration reduce to the two partial differential equations and the related jump conditions 
\begin{subequations}
\begin{align}
    \label{eq:maxwell_b}
    \div \ve{b} &= 0 \text{ with } \llbracket\ve{b}\rrbracket \cdot \ve{n} = 0 \text{ on } \mathcal{S} \text{ and} \\
    \label{eq:maxwell_h}
    \rot \ve{h} &= \ve{0} \text{ with } \llbracket\ve{h}\rrbracket \times \ve{n} = \ve{0} \text{ on } \mathcal{S}
\end{align}
\end{subequations}
with the magnetic induction $\ve{b} \in \mathcal{L}_1$ and the magnetic field $\ve{h} \in \mathcal{L}_1$.
The jump operator $\llbracket(\circ)\rrbracket = (\circ)^+ - (\circ)^-$ describes the change of a variable $(\circ)$ over a material surface $\mathcal{S} \subseteq \mathcal{B}$ with the normal unit vector $\ve{n} \in \mathcal{N}$ from subdomain $\mathcal{B}^- \subseteq \mathcal{B}$ to $\mathcal{B}^+ \subseteq \mathcal{B}$.
In the case of MREs, the subdomains can be the matrix and the particles, or the MRE and the surrounding vacuum, respectively.
The two introduced magnetic variables are connected by the equation
\begin{equation}
    \label{eq:b_h_m}
    \ve{b} = \mu_0 (\ve{h} + \ve{m}) ,
\end{equation}
where $\mu_0 = 4\pi\cdot10^{-7} \,\si{\newton\per\ampere\squared}$ is the vacuum permeability and $\ve{m} \in \mathcal{L}_1$ is the magnetization.
By using the pull-back operations \cite{Dorfmann04,Kankanala04}
\begin{equation}
    \label{eq:pull-back}
    \ve{B} = \ve{b} \cdot \op{cof}(\te{F}), \ \
    \ve{H} = \ve{h} \cdot \te{F}, \ \ 
    \ve{M} = \ve{m} \cdot \te{F} \ \  \text{and} \ \ 
    \ve{N} = \frac{\ve{n} \cdot \te{F}}{\left|\ve{n} \cdot \te{F}\right|} ,
\end{equation}
the above-mentioned equations can be re-written in the reference configuration as
\begin{subequations}
\label{eq:Maxwell_HB}
\begin{align}
        \label{eq:maxwell_B}
        \Div \ve{B} &= 0 \text{ with } \llbracket\ve{B}\rrbracket \cdot \ve{N} = 0 \text{ on } \mathcal{S}_0 , \\
        \label{eq:maxwell_H}
        \Rot \ve{H} &= \ve{0} \text{ with } \llbracket\ve{H}\rrbracket \times \ve{N} = \ve{0} \text{ on } \mathcal{S}_0 \text{ and }
\end{align}
\end{subequations}
\begin{equation}
    \label{eq:B_H_M}
    \ve{B} = \mu_0 J \te{C}^{-1} \cdot (\ve{H} + \ve{M}) .
\end{equation}
Here, 
$\ve{B} \in \mathcal{L}_1$, $\ve{H} \in \mathcal{L}_1$, $\ve{M} \in \mathcal{L}_1$, $\ve{N} \in \mathcal{N}$ and $\mathcal{S}_0 \subseteq \mathcal{B}_0$ are the magnetic induction, magnetic field, magnetization, normal unit vector and material surface in the reference configuration, respectively.

Additionally to mechanically induced loads, the ponderomotive force density $\ve{f}^\text{pon} = \ve{m} \cdot \grad \ve{b} \in \mathcal{L}_1$, the ponderomotive torque density $\ve{c}^\text{pon} = \ve{m} \times \ve{b} \in \mathcal{L}_1$ and the ponderomotive power density $p^\text{pon} = \ve{f}^\text{pon} \cdot \ve{v} - \ve{m} \cdot \dot{\ve{b}} \in \mathbb{R}$ act in the magneto-mechanical case.
The specific formulation of these ponderomotive quantities is based on the field-matter interaction model of de Groot and Suttorp \cite{degroot69}, though alternative approaches have also been proposed \cite{Pao78}.
To consider both the ponderomotive force and torque density, the ponderomotive stress tensor $\te{\upsigma}^\text{pon} \in \mathcal{L}_2$ is introduced such that it satisfies the equations
\begin{subequations}
\begin{align}
    \label{eq:f_c_pon_sig_pon}
    \div (\te{\upsigma}^\text{pon})^\text{T} &= \ve{f}^\text{pon} \text{ and } \\
    \op{skw}(\te{\upsigma}^\text{pon}) &= - \frac{1}{2} \mathfrak{e} \cdot \ve{c}^\text{pon} ,
\end{align}
\end{subequations}
resulting in the formulation 
\begin{equation}
    \label{eq:sig_pon}
    \te{\upsigma}^\text{pon} = \frac{1}{\mu_0}\left( \ve{b} \otimes \ve{b} - \frac{1}{2}\left|\ve{b}\right|^2 \te{1} \right) + (\ve{b} \cdot \ve{m}) \te{1} - \ve{m} \otimes \ve{b}
\end{equation}
for the ponderomotive stress tensor \cite{degroot69,Kalina20-1}.
The jump of the ponderomotive stress tensor over the material surface $\mathcal{S}$ is given by the ponderomotive stress vector $\ve{t}^\text{pon} = \jump{\te{\upsigma}^\text{pon}} \cdot \ve{n}$.
After defining the total stress tensor 
\begin{equation}
    \label{eq:sig_tot}
    \te{\upsigma}^\text{tot} = \te{\upsigma} + \te{\upsigma}^\text{pon} \in \mathscr{Sym}
\end{equation}
with the mechanical stress tensor $\te{\upsigma}$ and by using the 
balance of mass $\rho_0 = J \rho$ with the mass densities $\rho_0$ and $\rho$ in the reference and the current configuration, respectively, the balance of linear and angular momentum in the absence of external mechanical forces can be expressed as
\begin{subequations}
    \begin{align}
        \label{eq:momentum_balance_sig}
        \div \te{\upsigma}^\text{tot} &= \ve{0} \quad \text{and} \\
        \op{skw}(\te{\upsigma}^\text{tot}) &= \te{0}
    \end{align}
\end{subequations}
with the jump condition $\llbracket \te{\upsigma}^\text{tot} \rrbracket \cdot \ve{n} = \ve{0} \text{ on } \mathcal{S}$.
Note, that the tensors $\te{\upsigma}^\text{pon}$ and $\te{\upsigma}$ are not symmetric in general.
By using the balance of energy and entropy while neglecting thermal effects, the Clausius-Duhem inequality
\begin{equation}
    \label{eq:CDU}
    -J^{-1} \dot{\Psi} + 
    \left( \te{\upsigma} + \frac{1}{2 \mu_0} \left|\ve{b}\right|^2 \te{1} \right) :
    \left( \dot{\te{F}} \cdot \te{F}^{-1} \right)^\text{\!T} + 
    \ve{h} \cdot \dot{\ve{b}} \ge 0 
\end{equation}
is formulated in the current configuration.
The total Helmholtz free energy density, denoted by $\Psi \in \mathbb{R}_{\ge 0}$, is referred to below
simply as free energy.
By using the pull-back operations defined in Equation~\eqref{eq:pull-back} and the total first Piola-Kirchhoff stress tensor $\Ptot = \sigtot \cdot \op{cof}(\te{F}) \in \mathcal{L}_2$, the balance of linear and angular momentum and the Clausius-Duhem inequality in the reference configuration are given by
\begin{subequations}
    \begin{align}
        \label{eq:momentum_balance_P}
        \Div (\Ptot)^\text{T} &= \ve{0} , \\
        \op{skw}(\Ptot \cdot \te{F}^\text{T}) &= \te{0} \quad \text{and}
    \end{align}
\end{subequations}
\begin{equation}
    \label{eq:CDU_euler}
    - \dot{\Psi} + \Ptot : \dot{\te{F}}{}^\text{T} + \ve{H} \cdot \dot{\ve{B}} \ge 0 
\end{equation}
with the jump condition $\llbracket \Ptot \rrbracket \cdot \ve{N} = \ve{0} \text{ on } \mathcal{S}_0$.

\subsection{Constitutive equations}\label{sec:Constitutive_equations}

In the following, we assume magnetically soft and elastic inclusions and a non-magnetizable and elastic matrix, which results in a path-independent behavior, i.e., $\Psi = \Psi(\te{F}, \ve{B})$.
According to the procedure of Coleman and Noll \cite{Coleman1963}, Inequation~\eqref{eq:CDU_euler} holds true if
\begin{equation}
    \label{eq:Ptot_Psi_F_H_Psi_B}
        \Ptot = \pd{\Psi}{\te{F}} 
        \quad \text {and} \quad 
        \ve{H} = \pd{\Psi}{\ve{B}} .
\end{equation}
Therefore, within magneto-hyperelasticity, the formulation of the free energy $\Psi(\te{F},\ve{B})$ fully describes the material behavior and is sufficient for the calculation of all magneto-mechanical quantities introduced in Section~\ref{sec:Maxwell_and_balance_equations} for a given $\te{F}$ and $\ve{B}$.
In the following, we outline several conditions on the free energy, most of which were summarized by Linden~et~al. \cite{Linden23} and Kalina~et~al. \cite{Kalina24}.
The material objectivity
\begin{equation}
    \label{eq:objectivity}
    \Psi(\te{F},\ve{B}) = \Psi(\te{Q} \cdot \te{F}, \ve{B}) 
    \ \forall \te{F} \in \mathcal{GL}^+(3), \ve{B} \in \mathcal{L}_1, \te{Q} \in \mathcal{SO}(3) , 
\end{equation}
ensures independence of the material behavior from the chosen frame of reference.
Material symmetry
\begin{equation}
        \label{eq:mat_symmetry}
    \Psi(\te{F},\ve{B}) = \Psi(\te{F} \cdot \te{Q}^\text{T}, \op{det}(\te{Q}) \te{Q} \cdot \ve{B}) 
    \ \forall \te{F} \in \mathcal{GL}^+(3), \ve{B} \in \mathcal{L}_1, \te{Q} \in \mathcal{G} \subseteq \mathcal{O}(3) 
\end{equation}
requires invariance with respect to rotations defined by the symmetry group $\mathcal{G}$, which depends on the anisotropy class of the material.
According to the condition of time reversal \cite{Eringen90}
\begin{equation}
    \label{eq:time_reversal}
    \Psi(\te{F}, \ve{B}) = \Psi(\te{F}, -\ve{B})
    \quad\text{for}\quad t \to -t ,
\end{equation}
the Maxwell and balance equations remain invariant under time reversal.
Under an infinitely large volumetric compression, expansion, or magnetic field, the free energy is expected to approach infinity, which is called the volumetric growth condition
\begin{equation}
    \label{eq:coercivity_J}
    \lim_{J \to 0} \Psi(\te{F},\ve{B}) = \lim_{J \to \infty} \Psi(\te{F},\ve{B}) = \infty .
\end{equation}
and magnetic growth condition
\begin{equation}
    \label{eq:magnetic_growth_condition}
    \lim_{|\ve{B}| \to \infty} \Psi(\te{F},\ve{B}) = \infty ,
\end{equation}
respectively.
In contrast to the free energy, the magnetization of ferromagnetic materials, which are commonly used as particles in MREs, shows the saturation behavior
\begin{equation}
    \label{eq:magnetic_saturation}
    \lim_{|\ve{B}| \to \infty} \norm{\ve{m}} = J^{-1} m_\text{s} 
    \quad\text{and}\quad
    \norm{\ve{m}} < J^{-1} m_\text{s} \ \forall \ve{m} \in \mathcal{L}_1 
\end{equation}
with the saturation magnetization $m_\text{s}$.
For composite materials with magnetizable particles, the upper bound of the magnetization $m_\text{s,MRE} = \phi m_\text{s}$ can be obtained by multiplying the saturation magnetization of the particles $m_\text{s}$ with their volume fraction $\phi \in \left(0,1\right)$ \cite{Danas12,Danas17}.
Furthermore, when considering only deformation-free states, ferromagnetic materials typically exhibit a magnetization that increases with an increasing magnetic field $\ve{b} = b \ve{n}$ with magnitude $b$ along a given direction $\ve{n}$ and follows a concave curve \cite{Devi21}, i.e.,
\begin{equation}
    \label{eq:magnetic_monotonic_concave}
\pd{\norm{\ve{m}}}{b} > 0 
\ \ \text{and} \ \
\frac{\partial^2 \norm{\ve{m}}}{\partial b^2} < 0
\ \ \forall b \in \mathbb{R}_{>0},\ \forall \ve{n} \in \mathcal{N},\ \te{F} = \te{1} .
\end{equation}
In the undeformed and/or purely mechanical case, the free energy, stress and magnetic field vanish, respectively, i.e.,
\begin{subequations}\label{eq:initial_zero}
\begin{align}
    \Psi(\te{F} = \te{1}, \ve{B} = \ve{0}) &= 0 \label{eq:psi_zero},\\
    \Ptot(\te{F} = \te{1}, \ve{B} = \ve{0}) &= \te{0} \label{eq:ptot_zero} \text{ and}\\
    \ve{H}(\te{F}, \ve{B} = \ve{0}) &= \ve{0} \label{eq:H_zero}.
\end{align}
\end{subequations}
The free energy of a material exposed to arbitrary loads is always non-negative, i.e.,
\begin{equation}
    \label{eq:psi_positive}
    \Psi(\te{F},\ve{B}) \ge 0 \ \forall \te{F} \in \mathcal{GL}^+(3), \ve{B} \in \mathcal{L}_1 .
\end{equation}
The specific equations for the microscale free energy of the constituents are given in Section~\ref{sec:Microscale_material_models}, while a PANN for the macroscale free energy of the composite material is introduced in Section~\ref{sec:Model_formulation}.

\subsection{Scale transition scheme}\label{sec:Scale_transition_scheme}

As mentioned in Section~\ref{Introduction}, it is infeasible to solve the microscopic BVP for real life macroscopic samples. 
Consequently, a distinction is made between the microscale and macroscale, with both being linked through a scale transition scheme.
At the microscale, we consider the heterogeneity of the material, which is characterized by the characteristic length $\mathscr{l} \in \mathbb{R}_{>0}$ of its constituents. 
At the macroscale, a homogeneous body with the characteristic length $\bar{\mathscr{l}} \in \mathbb{R}_{>0}$ is considered, and the microstructure is ignored.
Macroscopic variables are denoted as $(\bar{\circ})$.
The scale separation $\bar{\mathscr{l}} \gg \mathscr{l}$ has to hold true \cite{Schröder13}.
At the microscale, it is sufficient to consider a volume element that represents the macroscopic behavior of the material.
Such a volume element is called RVE and can be used to determine macroscopic variables by averaging.
For the RVE to be representative, these macroscopic variables must remain unchanged when the RVE's position is varied or the size is increased.
In statistical microstructures, the RVE must contain a sufficient amount of inclusions and has to be much smaller than the sample, i.e., $\mathscr{l} \ll \mathscr{l}^\text{RVE} \ll \bar{\mathscr{l}}$ \cite{Gross11}.
By volume averaging all microscale variables over the RVE in its reference configuration $\mathcal{B}_0^\text{RVE}$ with the volume $V^\text{RVE}$ using the averaging operator
\begin{equation}
    \ho{\circ}_0 = \frac{1}{V^\text{RVE}} \int_{\mathcal{B}_0^\text{RVE}} \circ \ \dd V ,
\end{equation}
inconsistencies can arise, as all equations introduced in the previous sections are valid for both the microscale and macroscale \cite{Schröder13}. 
Therefore, we define only the variables 
\begin{equation}
    \label{eq:homogenization_F_P_B_H}
    \bar{\te{F}} = \ho{\te{F}}_0 \  , 
    \  \bar{\te{P}}^\text{tot} = \ho{\te{P}^\text{tot}}_0 \  , 
    \  \ve{\bar{B}} = \ho{\ve{B}}_0 \  \text{and}
    \  \ve{\bar{H}} = \ho{\ve{H}}_0
\end{equation}
in this manner, with all other macroscopic quantities calculated based on them \cite{Chatzigeorgiou14}.
Fields on the microscale can be additively split into a macroscopic component and a fluctuation, denoted by $(\tilde{\circ})$.
For a given macroscale deformation gradient and magnetic induction, $\te{F}$ and $\ve{B}$ can therefore be expressed as
\begin{equation}
    \label{eq:F_B_split}
    \te{F}(\tilde{\ve{u}}) = \bar{\te{F}} + (\Grad \tilde{\ve{u}})^\text{T} 
    \quad \text{ and } \quad
    \ve{B}(\tilde{\ve{A}}) = \ve{\bar{B}} + \Rot \tilde{\ve{A}}
\end{equation}
with the fluctuations of the displacement $\tilde{\ve{u}} \in \mathcal{L}_1$ and the magnetic vector potential $\tilde{\ve{A}} \in \mathcal{L}_1$ with $\ve{N} \times \jump{\tilde{\ve{A}}} = \ve{0}$ on $\mathcal{S}_0$.
By assuming constant $\bar{\te{F}}$ and $\bar{\ve{B}}$ over the RVE,
\begin{equation}
    \label{eq:u_A_macro}
    \bar{\ve{u}} = (\bar{\te{F}} - \te{1}) \cdot \ve{X} 
    \quad \text{and} \quad
    \bar{\ve{A}} = \frac{1}{2} \ve{\bar{B}} \times \ve{X}
\end{equation}
can be obtained.
According to the Hill-Mandel condition \cite{Hill63,Chatzigeorgiou14}, given by
\begin{equation}
    \label{eq:hill-mandel}
    \ho{\te{P}^\text{tot} : \dot{\te{F}}{}^\text{T} + \ve{H} \cdot \dot{\ve{B}}}_0
    = \bar{\te{P}}^\text{tot} : \dot{\bar{\te{F}}}^\text{T} + \ve{\bar{H}} \cdot \dot{\bar{\ve{B}}}, 
\end{equation}
the macroscopic and averaged microscopic powers are equivalent.
There are several boundary conditions that can satisfy this requirement.
The ones chosen here are periodic boundary conditions \cite{Zabihyan18,Zabihyan20}, which can be formulated as
\begin{equation}
    \begin{aligned}
        \label{eq:periodic_bc}
            \periodplus{\tilde{\ve{u}}} &= \ve{0} \quad, 
            &\periodplus{\tilde{\ve{A}}} \times \ve{N} &= \ve{0} , \\ 
            \periodplus{\te{P}^\text{tot}} \cdot \ve{N} &= \ve{0} \quad \text{and} \
            &\periodplus{\ve{H}} \times \ve{N} &= \ve{0} .
    \end{aligned}
\end{equation}
Therein, we set the quantities $(\circ)^-$ and $(\circ)^+$ at corresponding points on opposite sides $\ve{X}^- \in \partial\mathcal{B}_0^{\text{RVE},-}$ and $\ve{X}^+ \in \partial\mathcal{B}_0^{\text{RVE},+}$ of the RVE into relation using the jump operator $\periodplus{(\circ)} = (\circ)^+ - (\circ)^-$.

\subsection{Variational formulation}\label{sec:Variational_formulation}

In this section, we outline the variational formulation of the microscale FE problem with the boundary conditions in Equation~\eqref{eq:periodic_bc}, which is used in Section~\ref{sec:Microscale_Finite_Element_simulations}. 
The same approach is used in Section~\ref{sec:Macroscale_magnetostrictive_effect} for macroscale FE simulations with $\bar{\te{F}}$ and $\bar{\ve{B}}$ from Equation~\eqref{eq:F_bar_B_bar_macro}, the primary variables $\bar{\ve{u}}, \bar{\ve{A}}, \theta, p$, and the boundary conditions from Equation~\eqref{eq:macro_bc}.

To solve the Maxwell equation \eqref{eq:maxwell_B}, the fluctuation of the magnetic vector potential $\tilde{\ve{A}}$ is defined according to Equation~\eqref{eq:F_B_split}.
Multiple solutions have been proposed to ensure the well-posedness of the problem \cite{Gebhart24,Biro89,ZaglmayrDiss}.
Here, a regularization approach \cite{Reitzinger02,ZaglmayrDiss} is chosen, where the term
\begin{equation}
    \label{eq:Pi_regularization}
    \Pi^\text{reg}(\tilde{\ve{A}}) = \int_{\mathcal{B}_0} \frac{\kappa}{2}\norm{\tilde{\ve{A}}}^2 \ \dd V
\end{equation}
is added to the potential with the small regularization parameter $\kappa \in \mathbb{R}_{>0}$.
To describe deformations, the fluctuation of the displacement $\tilde{\ve{u}}$ is used as the second primary field variable.
Those two primary field variables are sufficient to define the magneto-hyperelastic problem, but the resulting formulation leads to locking for quasi-incompressible materials, such as silicone rubber. 
Therefore, the dilatation $\theta$ and pressure $p$ are added to the list of primary field variables \cite{Simo85,Okada09,Polukhov18}, which results in the four-field formulation as used by Polukhov~et~al. \cite{Polukhov20}, which is presented in the following.
The free energy
\begin{equation}
    \Psi(\te{F}, \ve{B}, \theta, p) = 
    \Psi^\text{mech}(\te{F}^\text{iso}, \theta) + 
    \Psi^\text{coup}(\te{F},\ve{B}) +
    p (J - \theta) ,
\end{equation}
consists of a mechanical and a coupled magneto-mechanical free energy and a term enforcing the equality of $J = \op{det}(\te{F})$ and $\theta$ with the Lagrange multiplier $p$.
The isochoric deformation gradient is given as $\te{F}^\text{iso}(\tilde{\ve{u}}) = \op{det}\left(\te{F}(\tilde{\ve{u}})\right)^{-\frac{1}{3}} \te{F}(\tilde{\ve{u}}) \in \mathscr{Uni}$, with $\te{F}(\tilde{\ve{u}})$ and $\ve{B}(\tilde{\ve{A}})$ being defined according to Equation~\eqref{eq:F_B_split} for given $\bar{\te{F}}$ and $\bar{\ve{B}}$.
The associated potential when considering the regularization in Equation~\eqref{eq:Pi_regularization} is defined as
\begin{equation}
    \label{eq:Pi_micro}
    \Pi(\tilde{\ve{u}}, \tilde{\ve{A}}, \theta, p) = 
    \int_{\mathcal{B}_0} \Psi(\te{F}, \ve{B}, \theta, p) \ \dd V 
    + \Pi^\text{reg}(\tilde{\ve{A}}) .
\end{equation}
The optimality condition for the variational problem is given by the first variation $\delta \Pi = 0$, which leads to the Euler-Lagrange equations 
\begin{subequations}
    \label{eq:Euler_Lagrange}
    \begin{align}
        \Div\left(\pd{\Psi}{\te{F}}\right)^\text{\!\!T} &= \ve{0} \ \text{in} \ \mathcal{B}_0 , \label{eq:Euler_Lagrange_lin_momentum} \\
        \Rot\pd{\Psi}{\ve{B}} + \kappa \tilde{\ve{A}} &= \ve{0} \ \text{in} \ \mathcal{B}_0 , \label{eq:Euler_Lagrange_Ampere} \\
        \pd{\Psi}{\theta} - p &= 0 \  \text{in} \ \mathcal{B}_0 , \label{eq:Euler_Lagrange_p} \\
        J - \theta &= 0 \ \text{in} \ \mathcal{B}_0 \label{eq:Euler_Lagrange_theta} ,
    \end{align}
\end{subequations}
with the jump conditions $\jump{\pd{\Psi}{\te{F}}} \cdot \ve{N} = \ve{0}$ and $\jump{\pd{\Psi}{\ve{B}}} \times \ve{N}= \ve{0}$.
The Equations~\eqref{eq:Euler_Lagrange_lin_momentum} and \eqref{eq:Euler_Lagrange_Ampere} are equivalent to Equations~ \eqref{eq:momentum_balance_P} and \eqref{eq:maxwell_H} for small $\kappa$, respectively \cite{Polukhov20,Gebhart24}.

For further information on the extremum principle, the used function spaces, and the FE discretization, the reader is referred to Appendix~\ref{app:Finite_Element_discretization}.

\section{Data-driven multiscale scheme}\label{chap:Data-driven_multiscale_scheme}

\begin{figure*}[]
    \centering
    \centerline{\includegraphics[width=\textwidth]{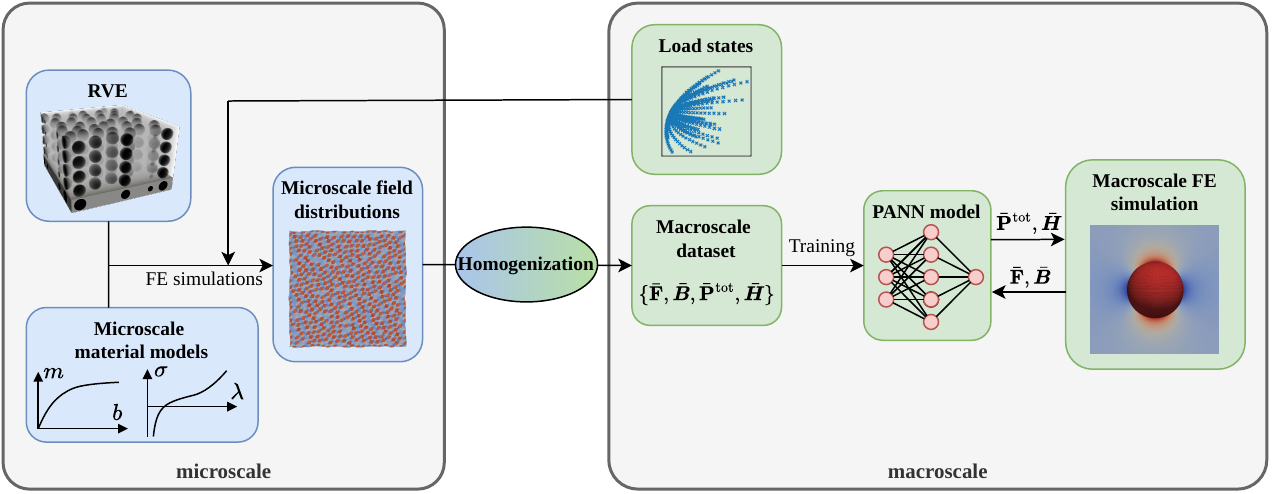}}
    \caption{Decoupled data-driven multiscale scheme: Macroscopic magneto-mechanical loads are applied to the RVE, and the resulting microscale fields are computed with the FEM from the known material behavior of the constituents. Homogenization yields a macroscale dataset for training the PANN model, which allows efficient macroscale FE simulations without explicitly solving microscopic BVPs.}
    \label{fig:multiscale_scheme}
\end{figure*}

As discussed in Section~\ref{sec:Modeling_approaches_for_MREs}, coupled multiscale schemes are too computationally expensive for the modeling of MREs.
Therefore, this work employs a decoupled data-driven multiscale scheme, illustrated in Figure~\ref{fig:multiscale_scheme}.
This approach relies on a surrogate model for the macroscale constitutive response, trained with homogenization data obtained from microscale simulations.
Starting on the microscale, the constitutive behavior of the components, i.e., the matrix and particles, is defined.
The computational domain on the microscale is an RVE, chosen such that its microstructure resembles that of a structured MRE under periodic boundary conditions.
A set of load states is then applied, and the corresponding macroscale response is obtained by solving the microscopic BVP and homogenizing the results.
This data is used to train a PANN model, enabling it to predict the macroscale constitutive behavior.
In the macroscale simulation, the constitutive response is provided directly by the PANN surrogate, different from a coupled multiscale scheme where a microscopic BVP must be solved for each occurring macroscale load state, which leads to a significantly reduced computation time.

In the following, the composite material is assumed to exhibit transverse isotropy with the preferred direction $\bar{\ve{S}} \in \mathcal{N}$.
Under this assumption, the condition of material symmetry can be fulfilled by using the structure tensor $\bar{\ve{S}} \otimes \bar{\ve{S}}$ and the isotropic tensor function 
$\bar{\Psi}(\bar{\te{F}}, \bar{\ve{B}}, \bar{\ve{S}}) = \bar{\Psi}(\bar{\te{F}} \cdot \te{Q}^\text{T}, \bar{\ve{B}} \cdot \te{Q}^\text{T} \det(\te{Q}), \te{Q} \cdot (\bar{\ve{S}} \otimes \bar{\ve{S}}) \cdot \te{Q}^\text{T}) \ \forall \te{Q} \in \mathcal{O}(3)$ 
\cite{Haupt2002,Itskov15,Kalina24_anisotropy}.
This condition is fulfilled if $\bar{\Psi}$ depends only on the invariants
\begin{equation}
    \label{eq:invariants_list}
    \begin{gathered}
        \bar{I}_1^\text{iso} = \op{tr}\bar{\te{C}}^\text{iso} ,  \quad
        \bar{I}_2^\text{iso} = \frac{1}{2}\left(\op{tr}^2\bar{\te{C}}^\text{iso} - \op{tr}(\bar{\te{C}}^\text{iso})^2\right) , \quad
        \bar{I}_3 = \bar{J}^2 ,  
        \bar{I}_4 = \bar{\ve{B}} \cdot \bar{\ve{B}} , \quad
        \bar{I}_5^\text{iso} = \bar{\ve{B}} \cdot \bar{\te{C}}^\text{iso} \cdot \bar{\ve{B}} , \quad \\
        \bar{I}_6^\text{iso} = \bar{\ve{B}} \cdot (\bar{\te{C}}^\text{iso})^2 \cdot \bar{\ve{B}} , \quad
        \bar{I}_7^\text{iso} = \bar{\ve{S}} \cdot \bar{\te{C}}^\text{iso} \cdot \bar{\ve{S}} , \quad
        \bar{I}_8^\text{iso} = \bar{\ve{S}} \cdot (\bar{\te{C}}^\text{iso})^2 \cdot \bar{\ve{S}} , \quad 
        \bar{I}_9 = (\bar{\ve{S}} \cdot \bar{\ve{B}})^2 \quad \text{and} \quad
        \bar{I}_{10}^\text{iso} = (\bar{\ve{S}} \cdot \bar{\te{C}}^\text{iso} \cdot \bar{\ve{B}})^2 ,
    \end{gathered}
\end{equation} which form a functional basis to model macroscopic transversely isotropic magneto-hyperelasticity \cite{Bustamante07Diss}.\footnote{Other invariants are possible as well. Here, the invariants forming an integrity basis for the anisotropy type with respect to the group $C_{\infty v}$ chosen by Bustamante \cite{Bustamante07Diss} are adapted such that they only contain either volumetric or isochoric contributions, to make the model compatible with the FE formulation in Section~\ref{sec:Macroscale_simulation}. Additionally, $\bar{I}_9$ and $\bar{I}^\text{iso}_{10}$ are squared, to ensure $\pd{\bar{\Psi}(\underline{\bar{I}}(\bar{\te{F}},\bar{\ve{B}},\bar{\ve{S}}))}{\bar{\ve{B}}} \big|_{\bar{\ve{B}}=\ve{0}} = \ve{0}$ and $\underline{\bar{I}}(\bar{\te{F}},\bar{\ve{B}},\bar{\ve{S}}) = \underline{\bar{I}}(\bar{\te{F}},-\bar{\ve{B}},\bar{\ve{S}})$.}
The invariants containing a mechanical contribution depend either on $\bar{J}$ or on $\bar{\te{C}}^\text{iso} = \bar{J}^{-\frac{2}{3}} \bar{\te{C}} \in \mathscr{Uni} \cap \mathscr{Sym}$.
By formulating $\bar{\Psi}$ with invariants of $\bar{\te{C}}$, $\bar{\ve{B}}$ and $\bar{\ve{S}}$, the material objectivity, material symmetry and the symmetry of the total stress tensor are satisfied \cite{Linden23}.
Under time reversal, the magnetic induction changes sign, whereas the deformation gradient remains unaffected, i.e., $\bar{\ve{B}} \to -\bar{\ve{B}}$ and $\bar{\te{F}} \to \bar{\te{F}}$ for $t \to -t$ \cite{Eringen90}.
Since $\bar{\Psi}$ and $\bar{\te{P}}^\text{tot}$ depend quadratically and $\bar{\ve{H}}$ depends linearly on $\bar{\ve{B}}$, the transformations $\bar{\Psi} \to \bar{\Psi}$, $\bar{\te{P}}^\text{tot} \to \bar{\te{P}}^\text{tot}$ and $\bar{\ve{H}} \to - \bar{\ve{H}}$ follow, which are in agreement to the principle of time reversal in Equation~\ref{eq:time_reversal}.

\subsection{Data generation}\label{sec:Data_generation_theory}

For the training of the model in Section~\ref{sec:Training_procedure}, data tuples $\{ {}^i \bar{\te{F}}, {}^i \bar{\ve{B}}, {}^i \bar{\te{P}}^\text{tot}, {}^i \bar{\ve{H}} \}$ are required.
For this reason, we create a set of macroscopic load states $\{ {}^i \bar{\te{F}}, {}^i \bar{\ve{B}} \}$.
The number of load states should be minimal to reduce computational cost, while still sufficient to provide adequate information for the training of the model. Therefore, a sampling approach is employed as described in Appendix~\ref{app:Data_sampling}.
Subsequently, we create an RVE with periodic boundary conditions, whose microstructure resembles that of the MRE to be modeled.
Given the microscale material laws, the FE formulation from Section~\ref{sec:Variational_formulation} is used to compute the resulting microscale field distributions under the prescribed loadings.
These fields are then averaged according to Section~\ref{sec:Scale_transition_scheme}, yielding the required macroscale response $\{ {}^i \bar{\te{P}}^\text{tot}, {}^i \bar{\ve{H}} \}$.

\subsection{Physics-augmented neural network macroscale model}\label{sec:Neural_network-based_macroscale_constitutive_model}

The material model introduced in this section is based on feedforward neural networks (FNNs), which are composed of multiple sequential layers, where the output of each layer serves as the input to the next.
The first and last layers are referred to as the input and output layers, respectively, while the intermediate ones are called hidden layers.
Each layer contains a set of parameters --- weights and biases --- which are adapted during training \cite{Kollmannsberger2021}.
A particular subclass of FNNs is the positive neural network (PNN), which guarantees non-negative outputs.
This behavior is enforced by constraining the weights and biases of the output layer to be non-negative and by employing only positive-valued activation functions in the final hidden layer.
The set of permitted weights and biases of a PNN is denoted with $\mathcal{PNN}$.
For an efficient training, the input and output of a PNN should be normalized, which can be achieved with non-trainable normalization layers.
They transform the input $i_\alpha$ and output $\mathfrak{o}_\alpha$ to
    \begin{equation}
        \label{eq:input_output_norm}
        \mathfrak{i}_\alpha = \frac{2 i_\alpha + i_\alpha^\text{max} + i_\alpha^\text{min}}{i_\alpha^\text{max} - i_\alpha^\text{min}}
    \quad\text{and}\quad
    o_\alpha = \mathfrak{o}_\alpha (o_\alpha^\text{max} - o_\alpha^\text{min}) + o_\alpha^\text{min} ,
\end{equation}
respectively, with $i_\alpha^\text{min}$, $i_\alpha^\text{max}$, $o_\alpha^\text{min}$ and $o_\alpha^\text{max}$ being the minimal and maximal input and output values from the training dataset, respectively \cite{Kalina24}.

\subsubsection{Model formulation}\label{sec:Model_formulation}

\begin{figure*}[]
    \centering
    \centerline{\includegraphics[width=\textwidth]{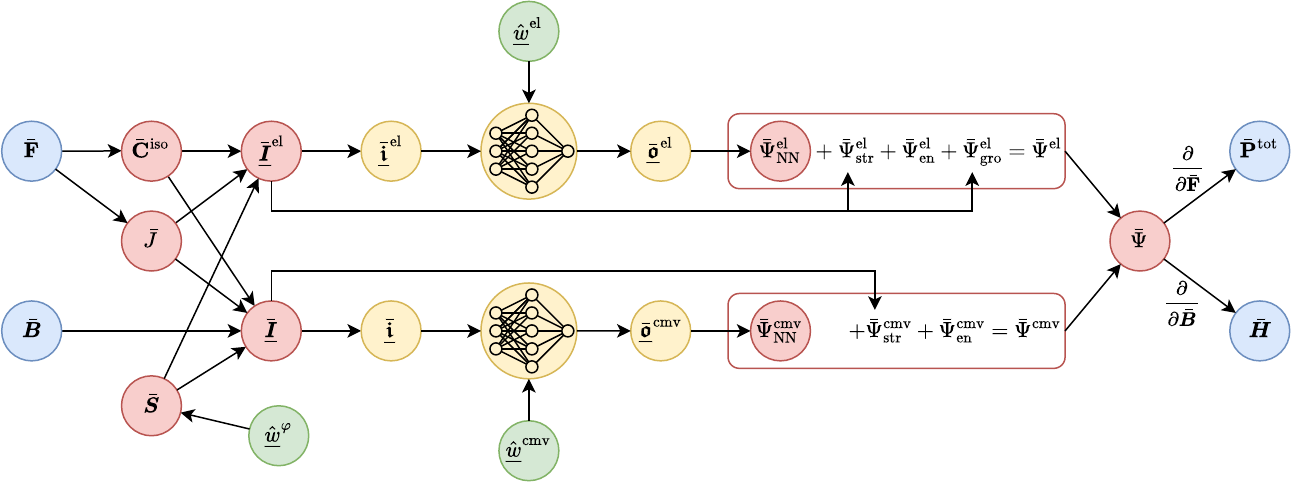}}
    \caption[]{Illustration of the macroscopic PANN model for transversely isotropic magneto-elasticity. The input and output of the model are displayed in blue, trainable variables in green, the neural network with its input and output normalization in yellow and all remaining variables in red. The layout was inspired by \cite{Linden23,Kalina24}.}
    \label{fig:diagram_model}
\end{figure*}

The invariants defined in Equation~\eqref{eq:invariants_list}, which are united in the invariant vector $\bar{\Ve{I}}(\bar{\te{F}}, \bar{\ve{B}}, \bar{\ve{S}}) = (\bar{I}_1^\text{iso}, \bar{I}_2^\text{iso}, \bar{I}_3, \bar{I}_4, \bar{I}_5^\text{iso}, \bar{I}_6^\text{iso}, \bar{I}_7^\text{iso}, \bar{I}_8^\text{iso}, \bar{I}_9, \bar{I}_{10}^\text{iso})$, are chosen as input of the model.
The purely mechanical invariants are combined in $\bar{\Ve{I}}^{\text{el}}(\bar{\te{F}}, \bar{\ve{S}}) = (\bar{I}_1^\text{iso},\bar{I}_2^\text{iso},\bar{I}_3,\bar{I}_7^\text{iso},\bar{I}_8^\text{iso})$.
As the preferred direction $\bar{\ve{S}}$ is constant for a given material and coordinate system and is learned as shown in Section~\ref{sec:Training_procedure}, it is a constant parameter of the function $\bar{\Psi}(\bar{\te{F}}, \bar{\ve{B}})$ during the prediction process.

In the following, the model described by Kalina~et~al. \cite{Kalina24} is extended to transverse isotropy.
This is based on a general approach for the construction of anisotropic neural network-based models using generalized structure tensors by Kalina~et~al. \cite{Kalina24_anisotropy}.
Emphasis is placed on satisfying the conditions described in Section~\ref{sec:Constitutive_equations}.
The free energy
\begin{equation}
    \label{eq:psi_model}
    \bar{\Psi}(\bar{\Ve{I}}) =
    \bar{\Psi}^\text{el}(\bar{\Ve{I}}^{\text{el}}) + 
    \bar{\Psi}^\text{cmv}(\bar{\Ve{I}})
\end{equation}
is split into a purely mechanical part $\bar{\Psi}^\text{el}$ and a part $\bar{\Psi}^\text{cmv}$ that includes the magneto-mechanical coupling, magnetic, as well as the vacuum energy.
Note that the Jacobian determinant $\bar{J} = \sqrt{\bar{I}_3}$ can be calculated from the invariants. 
The Equations~\eqref{eq:initial_zero} will be satisfied by fulfilling the sufficient conditions
\begin{subequations}
\begin{align}
    \bar{\Psi}^\text{el}(\bar{\te{F}} = \te{1}) = 0 
        \quad& \land \quad
    \bar{\Psi}^\text{cmv}(\bar{\te{F}} = \te{1}, \bar{\ve{B}} = \ve{0}) = 0 , \\
    \pd{\bar{\Psi}^\text{el}}{\bar{\te{F}}} \bigg|_{\bar{\te{F}}=\te{1}} = \te{0} 
        \quad& \land \quad
    \pd{\bar{\Psi}^\text{cmv}}{\bar{\te{F}}} \bigg|_{\bar{\te{F}}=\te{1}, \bar{\ve{B}}=\ve{0}} = \te{0} 
         \quad \text{and} \\
    \pd{\bar{\Psi}^\text{el}}{\bar{\ve{B}}} = \te{0} 
        \quad& \land \quad
    \pd{\bar{\Psi}^\text{cmv}}{\bar{\ve{B}}} \bigg|_{\bar{\ve{B}}=\ve{0}} = \te{0} ,
\end{align}
\end{subequations}
respectively.
The mechanical part
\begin{equation}
    \label{eq:psi_el_model}
    \bar{\Psi}^\text{el}(\bar{\Ve{I}}^{\text{el}}) =
    \bar{\Psi}^\text{el}_\text{NN}(\bar{\Ve{I}}^{\text{el}}) +
    \bar{\Psi}^\text{el}_\text{str}(\bar{J}, \bar{I}_7^\text{iso}) +
    \bar{\Psi}^\text{el}_\text{en} +
    \bar{\Psi}^\text{el}_\text{gro}(\bar{J})
\end{equation}
consists of the output of a PNN $\bar{\Psi}^\text{el}_\text{NN}(\bar{\Ve{I}}^{\text{el}}) \ge 0 \ \forall \bar{\Ve{I}}^\text{el} \in \mathbb{R}^{6}$ with the input and output normalization given in Equation~\eqref{eq:input_output_norm} and several additional terms.
To ensure $\pd{\bar{\Psi}^\text{el}}{\bar{\te{F}}} \Big|_{\bar{\te{F}}=\te{1}} = \te{0}$, the term
\begin{equation}
    \label{eq:psi_el_str}
    \bar{\Psi}^\text{el}_\text{str}(\bar{J}, \bar{I}_7^\text{iso}) =
    - \mathfrak{n}^{\text{el},\circledcirc} (\bar{J} - 1)
    - \mathfrak{n}^{\text{el},\parallel} (\bar{I}_7^\text{iso} - 1)
\end{equation}
with
    \begin{equation}
        \mathfrak{n}^{\text{el},\circledcirc} =
        2 \pd{\bar{\Psi}^\text{el}_\text{NN}}{\bar{I}_3}
        \Bigg|_{\bar{\te{F}}=\te{1}}
        \in \mathbb{R}
\quad\text{and}\quad
        \mathfrak{n}^{\text{el},\parallel} =
        \left(\pd{\bar{\Psi}^\text{el}_\text{NN}}{\bar{I}_7^\text{iso}} +
        2 \pd{\bar{\Psi}^\text{el}_\text{NN}}{\bar{I}_8^\text{iso}}\right) \!\! \Bigg|_{\bar{\te{F}}=\te{1}}
        \in \mathbb{R}
    \end{equation}
is added \cite{Kalina24_anisotropy}.\footnote{In contrast to \cite{Kalina24_anisotropy}, no terms for $\bar{I}_1^\text{iso}$ and $\bar{I}_2^\text{iso}$ are necessary, as $\pd{\bar{I}_1^\text{iso}}{\bar{\te{F}}} \Big|_{\bar{\te{F}}=\te{1}} = \pd{\bar{I}_2^\text{iso}}{\bar{\te{F}}} \Big|_{\bar{\te{F}}=\te{1}} = \te{0}$.}
We consider condition $\bar{\Psi}^\text{el}(\bar{\te{F}} = \te{1}) = 0$ by adding the negated output of the PNN in the undeformed state
\begin{equation}
    \label{eq:psi_el_en}
    \bar{\Psi}^\text{el}_\text{en} =
    - \bar{\Psi}^\text{el}_\text{NN}(\bar{\Ve{I}}^{\text{el}}) \big|_{\bar{\te{F}}=\te{1}} ,
\end{equation}
whereas the volumetric growth condition \eqref{eq:coercivity_J} is fulfilled for the mechanical part of the free energy with the term \footnote{According to \cite{Kalina24}, a value between $10^{-2}$ and $10^{-3}$ of the material's initial stiffness should be used for $\lambda_\text{gro}$.}
\begin{equation}
    \label{eq:psi_el_gro}
    \bar{\Psi}^\text{el}_\text{gro}(\bar{J}) =
    \lambda_{\text{gro}}(\bar{J} + \bar{J}^{-1} - 2)^2 .
\end{equation}
Note that the correction terms are chosen in such a way that they do not interfere with the other corrections, e.g., $\bar{\Psi}^\text{el}_\text{str}\big|_{\bar{\te{F}}=\te{1}} = 
0$.

Similarly to the mechanical part, the free energy component describing the coupled magneto-mechanical behavior
\begin{equation}
    \bar{\Psi}^\text{cmv}(\bar{\Ve{I}}) = 
    \bar{\Psi}^\text{cmv}_\text{NN}(\bar{\Ve{I}}) + 
    \bar{\Psi}^\text{cmv}_\text{str}(\bar{J}, \bar{I}_7^\text{iso}) + 
    \bar{\Psi}^\text{cmv}_\text{en}
\end{equation}
consists of the output of a PNN $\bar{\Psi}^\text{cmv}_\text{NN}(\bar{\Ve{I}}) \ge 0 \ \forall \bar{\Ve{I}} \in \mathbb{R}^{10}$ with input and output normalization and the terms 
\begin{equation}
    \label{eq:psi_cmv_str}
    \bar{\Psi}^\text{cmv}_\text{str}(\bar{J}, \bar{I}_7^\text{iso}) =
    - \mathfrak{n}^{\text{cmv},\circledcirc} (\bar{J} - 1)
    - \mathfrak{n}^{\text{cmv},\parallel} (\bar{I}_7^\text{iso} - 1)
\end{equation}
with
    \begin{equation}
        \mathfrak{n}^{\text{cmv},\circledcirc} =
        2 \pd{\bar{\Psi}^\text{cmv}_\text{NN}}{\bar{I}_3}
        \Bigg|_{\bar{\te{F}}=\te{1}, \bar{\ve{B}}=\ve{0}}
        \in \mathbb{R} ,
    \quad
        \mathfrak{n}^{\text{cmv},\parallel} =
        \left(\pd{\bar{\Psi}^\text{cmv}_\text{NN}}{\bar{I}_7^\text{iso}} +
        2 \pd{\bar{\Psi}^\text{cmv}_\text{NN}}{\bar{I}_8^\text{iso}}\right) \!\! \Bigg|_{\bar{\te{F}}=\te{1}, \bar{\ve{B}}=\ve{0}}
        \in \mathbb{R}
    \end{equation}
and
\begin{equation}
    \label{eq:psi_cmv_en}
    \bar{\Psi}^\text{cmv}_\text{en} =
    - \bar{\Psi}^\text{cmv}_\text{NN}(\bar{\Ve{I}}) \big|_{\bar{\te{F}}=\te{1}, \bar{\ve{B}}=\ve{0}} ,
\end{equation}
which ensure $\pd{\bar{\Psi}^\text{cmv}}{\bar{\te{F}}} \Big|_{\bar{\te{F}}=\te{1}, \bar{\ve{B}}=\te{0}} = \te{0}$ and $\bar{\Psi}^\text{cmv}(\bar{\te{F}} = \te{1}) = 0$, respectively.
The condition $\pd{\bar{\Psi}^\text{cmv}}{\bar{\ve{B}}} \Big|_{\bar{\ve{B}}=\te{0}} = \ve{0}$ is fulfilled without additional terms.
The entire model is summarized in Figure~\ref{fig:diagram_model}.

\begin{remark}
  In contrast to the microscale material models defined in Section~\ref{sec:Microscale_material_models}, the macroscale model introduced here does not utilize the additive split of the mechanical free energy $\bar{\Psi}(\bar{\te{F}}) = \bar{\Psi}^\text{iso}(\bar{\te{F}}^\text{iso}) + \bar{\Psi}^\text{vol}(\bar{J})$, because a study of the purely mechanical homogenization data obtained in Section~\ref{sec:Data_generation} has shown, that the hydrostatic pressure $\bar{p}^\text{hyd} = -\frac{1}{3}\op{tr}\bar{\te{\upsigma}} \in \mathbb{R}$ not only depends on $\bar{J}$, but also on additional isochoric invariants, thus making a model utilizing the split unable to accurately describe the mechanical behavior.
\end{remark}

\subsubsection{Training procedure}\label{sec:Training_procedure}

To enable the two-step training procedure suggested in \cite{Kalina24}, the dataset $\mathcal{D}$ consists of a purely mechanical and a coupled magneto-mechanical dataset. Both are split in a training dataset used for the model training and a test dataset used for the model validation, respectively, i.e., 
$\mathcal{D} = \mathcal{D}_\text{train}^\text{mech} \ \cup \ \mathcal{D}_\text{train}^\text{coup} \cup \mathcal{D}_\text{test}^\text{mech} \ \cup \ \mathcal{D}_\text{test}^\text{coup}$ with the respective numbers of data tuples in each dataset $n_\text{train}^\text{mech}$, $n_\text{train}^\text{coup}$, $n_\text{test}^\text{mech}$ and $n_\text{test}^\text{coup}$.
To allow its automatic detection during the training, we specify the preferred direction
\begin{equation}
    [ \bar{\ve{S}} ] =
    \begin{pmatrix}
        \sin{\varphi_1} \cos{\varphi_2} \\
        \sin{\varphi_1} \sin{\varphi_2} \\
        \cos{\varphi_1}
    \end{pmatrix} .
\end{equation}
depending on the trainable angles $\varphi_1$ and $\varphi_2$, which are combined in $\Ve{\mathscr{w}}^{\varphi} = \{\varphi_1, \varphi_2\} \in \mathcal{M}$ with $\mathcal{M} = \{\varphi_1 \in \left[0, \frac{\pi}{2}\right], \varphi_2 \in \left[0,2\pi\right]\}$ \cite{Kalina24_anisotropy}.
Before starting the training, the parameters of the normalization according to Equation~\eqref{eq:input_output_norm} are identified.

In the following, we present the training procedure according to Kalina~et~al.~\cite{Kalina24}, which is done in two steps.
First, a purely mechanical dataset $\mathcal{D}^\text{mech} = \{ {}^1\mathcal{T}^\text{mech}, {}^2\mathcal{T}^\text{mech}, \dots, {}^{n^\text{mech}}\mathcal{T}^\text{mech} \}$ with homogenization data tuples ${}^i\mathcal{T}^\text{mech} = ({}^i\bar{\te{F}}^\text{mech}, {}^i\bar{\te{\sig}}^\text{mech})^\text{RVE}$ and $\bar{\ve{B}} = \ve{0}$ is used to train the neural network $\bar{\Psi}_\text{NN}^\text{el}$.
This is done by finding the weights and biases
\begin{equation}
    (\hat{\Ve{\mathscr{w}}}^\text{el}, \hat{\Ve{\mathscr{w}}}^{\varphi,\text{el}}) = 
    \underset{\Ve{\mathscr{w}}^\text{el} \in \mathcal{PNN}, \Ve{\mathscr{w}}^{\varphi} \in \mathcal{M}}{\text{arg\,min }} 
    \mathcal{L}^\text{el}
\end{equation}
which minimize the loss
\begin{equation}
    \mathcal{L}^\text{el} = 
    \frac{1}{n_\text{train}^\text{mech}}
    \frac{1}{s_{\te{\upsigma}^\text{mech}}}
    \sum_{i=1}^{n_\text{train}^\text{mech}}
    \frob{{}^i\bar{\te{\upsigma}}^\text{mech} - \bar{\te{\upsigma}}^\text{el}({}^i\bar{\te{F}}^\text{mech},\Ve{\mathscr{w}}^\text{el}, \Ve{\mathscr{w}}^{\varphi})}^2 \\, 
\end{equation}
where $ s_{\te{\upsigma}^\text{mech}} =
\op{max}\left( \lVert {}^1\bar{\te{\upsigma}}^\text{mech} \rVert^2, 
               \lVert {}^2\bar{\te{\upsigma}}^\text{mech} \rVert^2, \dots, 
               \lVert {}^{n_\text{train}^\text{mech}}\bar{\te{\upsigma}}^\text{mech} \rVert^2 \right) $
serves as a normalizing factor.
The value of $\bar{\te{\upsigma}}^\text{el}({}^i\bar{\te{F}}^\text{mech}, \Ve{\mathscr{w}}^\text{mech}, \Ve{\mathscr{w}}^{\varphi})$ is calculated by determining the elastic part of the free energy of the model given in Equation~\eqref{eq:psi_el_model} and by applying $\bar{\te{P}}^\text{el} = \pd{\bar{\Psi}^\text{el}}{\bar{\te{F}}}$ and $\bar{\te{P}}^\text{el} = \bar{\te{\upsigma}}^\text{el} \cdot \op{cof}(\bar{\te{F}})$.

In the second training step, a coupled dataset $\mathcal{D}^\text{coup} = \{ {}^1\mathcal{T}^\text{coup}, {}^2\mathcal{T}^\text{coup}, \dots, {}^{n^\text{coup}}\mathcal{T}^\text{coup} \}$ with the homogenization data tuples
${}^i\mathcal{T}^\text{coup} = ({}^i\bar{\te{F}}^\text{coup}, {}^i\bar{\ve{B}}^\text{coup}, {}^i\bar{\te{\upsigma}}^\text{coup}, {}^i\bar{\ve{m}}^\text{coup})^\text{RVE}$ and $\bar{\ve{B}} \ne \ve{0}$ is used to train $\bar{\Psi}_\text{PNN}^\text{cmv}$, together with an additional dataset $\mathcal{D}^\text{add}$ that enforces conditions on the magnetization curve in a weak sense.
Using the purely mechanical neural network with the weights and biases $\hat{\Ve{\mathscr{w}}}^\text{el}$ determined in the first step, as well as initializing the weights $\Ve{\mathscr{w}}^{\varphi}$ as $\hat{\Ve{\mathscr{w}}}^{\varphi,\text{el}}$, the weights and biases 
\begin{equation}
    (\hat{\Ve{\mathscr{w}}}^\text{cmv}, \hat{\Ve{\mathscr{w}}}^{\varphi}) = 
    \underset{\Ve{\mathscr{w}}^\text{cmv} \in \mathcal{PNN}, \Ve{\mathscr{w}}^{\varphi} \in \mathcal{M}}{\text{arg\,min }}
    \mathcal{L}^\text{cmv} ,
\end{equation}
which minimize the loss 
\begin{equation}
    \mathcal{L}^\text{cmv} = w^{\te{\upsigma}} \mathcal{L}^{\te{\upsigma}} + w^{\ve{m}} \mathcal{L}^{\ve{m}} + w^\text{sat} \mathcal{L}^\text{sat} + w^\text{mon} \mathcal{L}^\text{mon} + w^\text{con} \mathcal{L}^\text{con} ,
\end{equation}
are identified.
The weighting factors $w^{\te{\upsigma}}$, $w^{\ve{m}}$, $w^\text{sat}$, $w^\text{mon}$ and $w^\text{con}$ allow to weight certain loss terms higher or lower than others.
The term
\begin{equation}
    \mathcal{L}^{\te{\upsigma}} = 
    \frac{1}{n_\text{train}^\text{coup}}
    \frac{1}{s_{\te{\upsigma}}}
    \sum_{i=1}^{n_\text{train}^\text{coup}}
    \frob{{}^i\bar{\te{\upsigma}}^\text{coup} - \bar{\te{\upsigma}}({}^i\bar{\te{F}}^\text{coup}, {}^i\bar{\ve{B}}^\text{coup}, \hat{\Ve{\mathscr{w}}}^{\text{el}}, \Ve{\mathscr{w}}^{\text{cmv}}, \Ve{\mathscr{w}}^{\varphi})}^2 \\
\end{equation}
with the normalization factor
$ s_{\te{\upsigma}} =
\op{max}\left( \lVert {}^1\bar{\te{\upsigma}}^\text{coup} \rVert^2, 
               \lVert {}^2\bar{\te{\upsigma}}^\text{coup} \rVert^2, \dots, 
               \lVert {}^{n_\text{train}^\text{coup}}\bar{\te{\upsigma}}^\text{coup} \rVert^2 \right) $
enforces equality between the predicted and reference stress.
We choose the mechanical stress tensor $\bar{\te{\sig}}$ instead of $\bar{\te{\sig}}^\text{tot}$, as it shows a far higher sensitivity than the total stress tensor \cite{Kalina21Diss}. Furthermore, the mechanical stress tensor influences the magnetostrictive and magnetorheological effect \cite{Kalina24}.
Likewise, the term
\begin{equation}
    \mathcal{L}^{\ve{m}} = 
    \frac{1}{n_\text{train}^\text{coup}}
    \frac{1}{s_{\ve{m}}}
    \sum_{i=1}^{n_\text{train}^\text{coup}}
    \left|{}^i\bar{\ve{m}}^\text{coup} - \bar{\ve{m}}({}^i\bar{\te{F}}^\text{coup}, {}^i\bar{\ve{B}}^\text{coup}, \hat{\Ve{\mathscr{w}}}^{\text{el}}, \Ve{\mathscr{w}}^{\text{cmv}}, \Ve{\mathscr{w}}^{\varphi})\right|^2 ,
\end{equation}
with 
$ s_{\ve{m}} =
\op{max}\left( \lVert {}^1\bar{\ve{m}}^\text{coup} \rVert^2, 
               \lVert {}^2\bar{\ve{m}}^\text{coup} \rVert^2, \dots, 
               \lVert {}^{n_\text{train}^\text{coup}}\bar{\ve{m}}^\text{coup} \rVert^2 \right) $
enforces the accurate prediction of the magnetization.
Both predictions $\bar{\te{\upsigma}}$ and $\bar{\ve{m}}$ are calculated from the free energy defined by the PANN model.

The loss terms $\mathcal{L}^\text{sat}$, $\mathcal{L}^\text{mon}$ and $\mathcal{L}^\text{con}$ enforce the boundedness, monotony, and concavity of the magnetization curve $\norm{\bar{\ve{b}}}(\norm{\bar{\ve{m}}})$ in a weak sense, respectively.
For this purpose, the model is evaluated for additional purely magnetic load states in the dataset $\mathcal{D}^\text{add} = \big\{ ^{1,1}\bar{\ve{B}}^\text{add} , \dots , ^{n^\text{add}_\phi,n^\text{add}_b}\bar{\ve{B}}^\text{add} \big\}$,
where 
\begin{equation}
    ^{i,j}\bar{\ve{B}}^\text{add} = {}^j b^\text{add} \left(\frac{\ve{e}_1 - (\ve{e}_1 \cdot \bar{\ve{S}}) \bar{\ve{S}}}{\norm{\ve{e}_1 - (\ve{e}_1 \cdot \bar{\ve{S}}) \bar{\ve{S}}}} \cos{{}^i \phi^\text{add}} + \bar{\ve{S}} \sin{{}^i \phi^\text{add}} \right) 
\end{equation} 
for angles $^{i}\phi^\text{add} = \angle(^{i,j}\bar{\ve{B}}^\text{add}, \bar{\ve{S}}) \in \left[0, \frac{\pi}{2} \right]$, assuming $\bar{\ve{S}} \ne \pm \ve{e}_1$. The magnitude of the magnetic induction ${}^j b^\text{add} \in \left[ b^\text{add}_\text{min}, b^\text{add}_\text{max} \right]$ is chosen in a range, in which the conditions should be enforced, with a step size of $\Delta b$.
In the following, we denote the prediction of the PANN model for the magnitude of the magnetization as $^{i,j}m = \norm{\bar{\ve{m}}(\bar{\te{F}}=\te{1}, {}^{i,j}\bar{\ve{B}}^\text{add}, \hat{\Ve{\mathscr{w}}}^{\text{el}}, \Ve{\mathscr{w}}^{\text{cmv}}, \Ve{\mathscr{w}}^{\varphi})}$.
The total number of additional load states is given by $n^\text{add} = n^\text{add}_\phi \cdot n^\text{add}_b$.

The upper bound of the magnetization is given by the saturation magnetization of the composite material $m_\text{s,MRE}$, as described in Section~\ref{sec:Constitutive_equations}.
This allows for defining the loss
\begin{equation}
    \label{eq:loss_sat}
    \mathcal{L}^\text{sat} = \frac{1}{n^\text{add} m^2_\text{s,MRE}} \sum_{i=1}^{n_\phi^\text{add}} \sum_{j=1}^{n_b^\text{add}} \op{ReLU} \left( ^{i,j}m - m_\text{s,MRE} \right)^2
\end{equation}
with $\op{ReLU}(x) = \max(0,x)$.
The condition of a monotonically increasing magnetization curve requires a positive first derivative, which is implemented using a finite difference as
\begin{equation}
    \label{eq:loss_mon}
    \mathcal{L}^\text{mon} = \frac{(b^\text{add}_\text{max})^2}{n^\text{add} m^2_\text{s,MRE}} \sum_{i=1}^{n_\phi^\text{add}} \sum_{j=1}^{n_b^\text{add}-1} \op{ReLU} \left( - \frac{\, ^{i,j+1}m - ^{i,j}m}{\Delta b} \right)^{\!2} .
\end{equation}
For a concave function, the second derivative is negative, which is enforced with the loss
\begin{equation}
    \label{eq:loss_con}
    \mathcal{L}^\text{con} = \frac{(b^\text{add}_\text{max})^4}{n^\text{add} m^2_\text{s,MRE}} \sum_{i=1}^{n_\phi^\text{add}} \sum_{j=1}^{n_b^\text{add}-2} \op{ReLU} \left( \frac{^{i,j+2}m - 2 \, ^{i,j+1}m + ^{i,j}m}{\Delta b^2} \right)^{\!2} .
\end{equation}
All conditions which are fulfilled by the PANN model are summarized in Table~\ref{tab:conditions_PANN}.

\begin{table}[]
    \begin{center}
        \small
            \caption{Overview of conditions with their corresponding equations, indicating whether the PANN model satisfies them strongly (\textcolor{myGreen}{\cmark}), weakly (\textcolor{myYellow}{(\cmark)}), or not at all (\textcolor{red}{\xmark}), together with the equation essential for the fulfillment of the condition.}
        \begin{tabular}{ l c c c }
        \hline
            Condition & in Equation & Fulfilled & using Equation \\ 
            \hline
            Thermodynamic consistency & \eqref{eq:CDU} & \textcolor{myGreen}{\cmark} & \eqref{eq:Ptot_Psi_F_H_Psi_B} \\
            Material objectivity & \eqref{eq:objectivity} & \textcolor{myGreen}{\cmark} & \eqref{eq:invariants_list}\\
            Material symmetry & \eqref{eq:mat_symmetry} & \textcolor{myGreen}{\cmark} & \eqref{eq:invariants_list} \\
            Symmetry of the total stress & \eqref{eq:momentum_balance_sig} & \textcolor{myGreen}{\cmark} & \eqref{eq:invariants_list} \\
            Time reversal & \eqref{eq:time_reversal} & \textcolor{myGreen}{\cmark} & \eqref{eq:invariants_list} \\ 
            Volumetric growth condition & \eqref{eq:coercivity_J} & \textcolor{myGreen}{\cmark} & \eqref{eq:psi_el_gro} \\
            Magnetic growth condition & \eqref{eq:magnetic_growth_condition} & \textcolor{red}{\xmark} & -- \\
            Magnetic saturation & \eqref{eq:magnetic_saturation} & \textcolor{red}{\xmark} & --  \\ 
            Boundedness of magnetization & \eqref{eq:magnetic_saturation} & \textcolor{myYellow}{(\cmark)} & \eqref{eq:loss_sat} \\ 
            Increasing magnetization curve & \eqref{eq:magnetic_monotonic_concave} & \textcolor{myYellow}{(\cmark)} & \eqref{eq:loss_mon} \\
            Concave magnetization curve & \eqref{eq:magnetic_monotonic_concave} & \textcolor{myYellow}{(\cmark)} & \eqref{eq:loss_con} \\
            Zero energy in unloaded state & \eqref{eq:psi_zero} & \textcolor{myGreen}{\cmark} & \eqref{eq:psi_el_en},\eqref{eq:psi_cmv_en} \\
            Stress-free unloaded state & \eqref{eq:ptot_zero} & \textcolor{myGreen}{\cmark} & \eqref{eq:psi_el_str},\eqref{eq:psi_cmv_str} \\
            Non-magnetized unloaded state & \eqref{eq:H_zero} & \textcolor{myGreen}{\cmark} & \eqref{eq:invariants_list} \\
            Non-negative free energy & \eqref{eq:psi_positive} & \textcolor{red}{\xmark} & -- \\
            \hline
        \end{tabular}
    \label{tab:conditions_PANN}
    \end{center}
\end{table}

\subsection{Macroscale simulation}\label{sec:Macroscale_simulation}

For the macroscale FE simulations of the magnetostrictive effect in Section~\ref{sec:Macroscale_magnetostrictive_effect}, we adapt the variational formulation from Section~\ref{sec:Variational_formulation}.
    The domain $\bar{\mathcal{B}} = \bar{\mathcal{B}}^\text{mat} \cup \bar{\mathcal{B}}^\text{vac}$ consists of the domain where the MRE is present, contained within the vacuum domain.
As primary variables, $\bar{\ve{u}}, \bar{\ve{A}}, \theta, p$ are selected with
\begin{equation}
    \label{eq:F_bar_B_bar_macro}
    \bar{\te{F}} = \te{1} + (\Grad \bar{\ve{u}})^\text{T} 
    \quad \text{and} \quad 
    \bar{\ve{B}} = \bar{\ve{B}}^\infty + \Rot \bar{\ve{A}} .
\end{equation}
Here, $\bar{\ve{B}}^\infty$ is a homogeneous magnetic far field.
The boundary conditions
\begin{equation}
    \begin{aligned}
        \label{eq:macro_bc}
            \bar{\ve{u}} &= \ve{0} \text{ on } \partial\mathcal{B}_0 \text{ and}\\
            \bar{\ve{A}} \times \bar{\ve{N}} &= \ve{0} \text{ on } \partial\mathcal{B}_0
    \end{aligned}
\end{equation}
enforce $\partial\mathcal{B} = \partial\mathcal{B}_0$ and $\bar{\ve{B}} = \bar{\ve{b}}$ on $\partial\mathcal{B}_0$.
The vacuum domain can be modeled using a highly compliant pseudo-vacuum, as described in Section~\ref{sec:Macroscale_magnetostrictive_effect} \cite{Gebhart24Diss}.

\begin{remark}
  If the geometry, the applied load, and the material behavior are symmetric with respect to all coordinate planes of a coordinate system, it is possible to reduce the computational costs by only calculating the field variables in the first octant.
  The applied magnetic induction $\bar{\ve{B}}^\infty$ and preferred material direction $\bar{\ve{S}}$ have to be either parallel or perpendicular to a coordinate axis.
  The boundary conditions in Equation~\eqref{eq:macro_bc} are adapted as follows for the example of $\bar{\ve{B}}^\infty \parallel \ve{e}_3$:
    \begin{equation}
        \begin{aligned}
            \label{eq:macro_bc_eight}
            \bar{\ve{u}} &= \ve{0} \text{ on } \partial\mathcal{B}_0^{x,+} \cup \partial\mathcal{B}_0^{y,+} \cup \partial\mathcal{B}_0^{z,+} , \\
            \ve{e}_1 \cdot \bar{\ve{u}} &= 0 \text{ on } \partial\mathcal{B}_0^{x,-} , \\
            \ve{e}_2 \cdot \bar{\ve{u}} &= 0 \text{ on } \partial\mathcal{B}_0^{y,-} , \\
            \ve{e}_3 \cdot \bar{\ve{u}} &= 0 \text{ on } \partial\mathcal{B}_0^{z,-} \text{ and }  \\
            \bar{\ve{N}} \times \bar{\ve{A}} &= \ve{0} \text{ on } \partial\mathcal{B}_0 \setminus \partial\mathcal{B}_0^{z,-} .
        \end{aligned}
    \end{equation}
    The notation \(\partial\mathcal{B}_0^{\alpha,\pm}\) represents the boundary surfaces whose outward normals are aligned with the positive or negative direction of the Cartesian coordinate axis \(\alpha \in \{x,y,z\}\).
\end{remark}

\section{Numerical examples}\label{chap:Numerical_examples}

We demonstrate the capabilities of the framework introduced in Section~\ref{chap:Data-driven_multiscale_scheme} using a structured MRE with preferred direction $\bar{\ve{S}}=\ve{e}_3$. For this purpose, we generate an RVE with an ideal chain structure and specify the microscale constitutive laws of its components. The RVE is then subjected to selected magneto-mechanical load states, and FEM-based homogenization is applied to compute the corresponding macroscale variables. Based on this dataset, we train a PANN model multiple times with randomly initialized parameters, and select the best-performing model. Its performance is assessed with respect to predictive accuracy for seen and unseen data, representation of elastic and magnetic anisotropy, and magnetic extrapolation capability. Subsequently, we employ the PANN model in a macroscale simulation to evaluate the magnetostriction of a sample. Finally, to further validate the approach, we compare predictions for load states at three representative points with reference solutions obtained from RVE simulations.

\subsection{Data generation}\label{sec:Data_generation}

In order to train the PANN model, macroscale data tuples are required, which are obtained in the following from the homogenization of microscale data from RVE simulations for sampled load cases and given microscale material models.

\subsubsection{Load sampling and RVE selection}\label{sec:Load_sampling}

\begin{figure}[]
    \centerline{\includegraphics[width=0.4\textwidth]{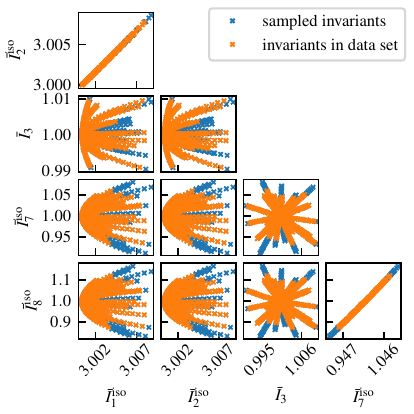}}
    \caption{Sampled and filtered mechanical invariants in the sampling ranges given in Table~\ref{tab:sampling_params} are shown in blue. The load cases for which convergence could be achieved during the FE microscale simulation are shown in orange and are used for the training and testing of the PANN model.\label{fig:sampled_invariants_mech}}
\end{figure}

\begin{figure*}[t]
    \centerline{\includegraphics[width=0.9\textwidth]{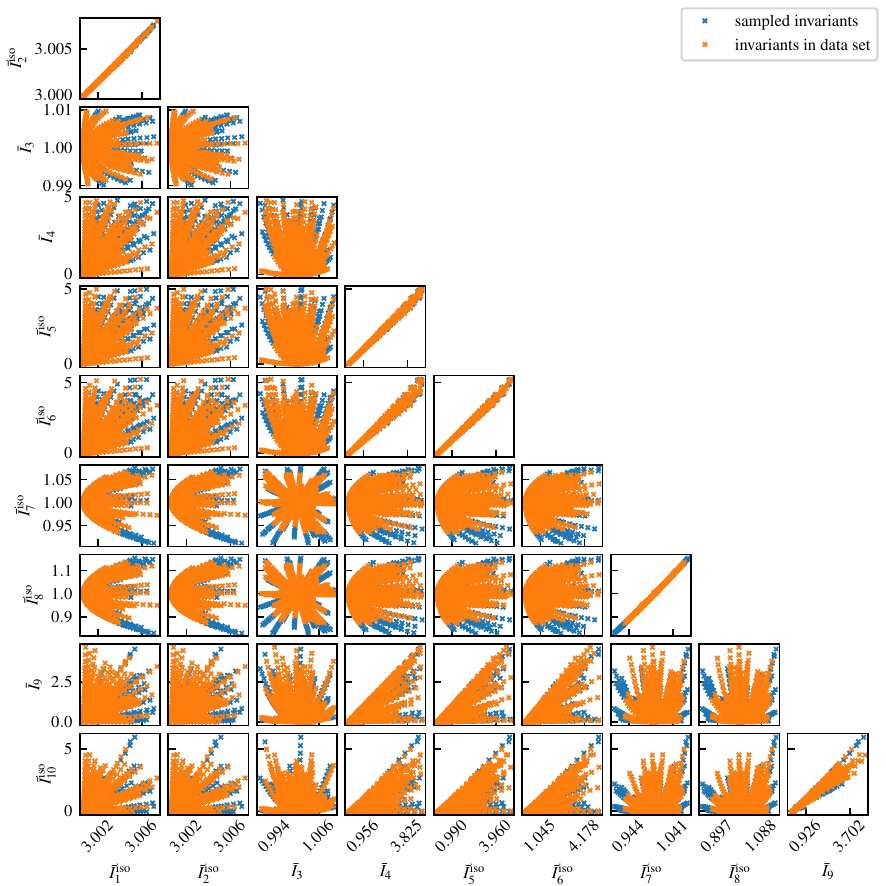}}
    \caption{Sampled and filtered coupled magneto-mechanical invariants in their SI base units in the sampling ranges given in Table~\ref{tab:sampling_params} are shown in blue. The load cases for which convergence could be achieved during the FE microscale simulation are shown in orange and are used for the training and testing of the PANN model.\label{fig:sampled_invariants_magmech}}
\end{figure*}

\begin{table*}[]
    \tiny
    \begin{center}
        \small
        \caption{Ranges for the sampling parameters, the total amount of remaining states after the sampling and total number of data tuples in the dataset. Each load path contains $M=20$ increments. The tolerance for filtering according to \cite{Kalina24} is $\epsilon_\text{tol} = 45\%$. The invariants corresponding to the sampled loads are shown in Figures~\ref{fig:sampled_invariants_mech} and \ref{fig:sampled_invariants_magmech}.}
        \begin{tabular}{ c c c c c c c c c c c }
        \hline
            type & $\bar{\lambda}_1$ & $\bar{\lambda}_2$ & $\bar{J}$ & $\theta_1$ & $\theta_2$ & $\phi_1$ & $\phi_2$ & $\bar{B}$ / \si{\tesla} & Filtered states & States in dataset \\ 
            \hline
            mechanical \begin{tabular}{@{}c@{}} min \\ max \end{tabular} &
            \begin{tabular}{@{}c@{}} 0.95 \\ 1.04 \end{tabular} &
            \begin{tabular}{@{}c@{}} 0.95 \\ 1.04 \end{tabular} &
            \begin{tabular}{@{}c@{}} 0.995 \\ 1.005 \end{tabular} &
            \begin{tabular}{@{}c@{}} 0 \\ 2$\pi$ \end{tabular} &
            \begin{tabular}{@{}c@{}} 0 \\ 2$\pi$ \end{tabular} &
            \begin{tabular}{@{}c@{}} -- \\ -- \end{tabular} &
            \begin{tabular}{@{}c@{}} -- \\ -- \end{tabular} &
            \begin{tabular}{@{}c@{}} -- \\ -- \end{tabular} &
            560 & 
            467 \\            
            \hline
            coupled \begin{tabular}{@{}c@{}} min \\ max \end{tabular} &
            \begin{tabular}{@{}c@{}} 0.95 \\ 1.04 \end{tabular} &
            \begin{tabular}{@{}c@{}} 0.95 \\ 1.04 \end{tabular} &
            \begin{tabular}{@{}c@{}} 0.995 \\ 1.005 \end{tabular} &
            \begin{tabular}{@{}c@{}} 0 \\ 2$\pi$ \end{tabular} &
            \begin{tabular}{@{}c@{}} 0 \\ 2$\pi$ \end{tabular} &
            \begin{tabular}{@{}c@{}} 0 \\ $\pi$/2 \end{tabular} &
            \begin{tabular}{@{}c@{}} 0 \\ 2$\pi$ \end{tabular} &
            \begin{tabular}{@{}c@{}} 0.5 \\ 2.2 \end{tabular} &
            1280 & 
            1125 \\
            \hline
        \end{tabular}
    \label{tab:sampling_params}
    \end{center}
\end{table*}

\begin{figure}[t]
    \centerline{\includegraphics[width=0.35\textwidth]{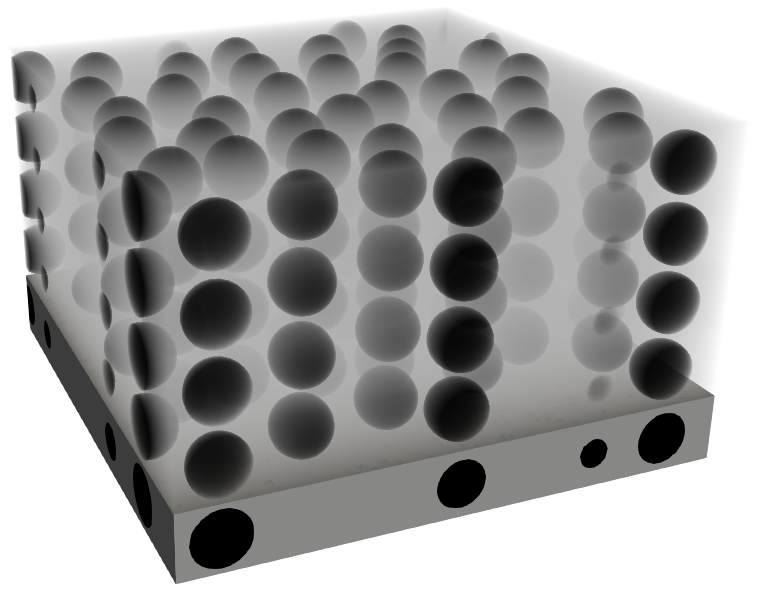}}
    \caption{Section of the RVE, periodically extended in $z$-direction to visualize the particles chains. The cross-section of the complete RVE is shown in Figure~\ref{fig:microscale_fields}.}
    \label{fig:RVE_3D_sw}
\end{figure}

For the sampling of load states, the approach described in Appendix~\ref{app:Data_sampling} is used. 
Under transverse isotropy, values for two principal stretches $\bar{\lambda}_1$ and $\bar{\lambda}_2$, the Jacobian determinant $\bar{J}$, two angles $\theta_1$ and $\theta_2$ describing the rotation of the deformation gradient, and two angles $\phi_1$ and $\phi_2$ describing the direction of the magnetic induction of magnitude $\bar{B}$ have to be sampled.
In Table~\ref{tab:sampling_params}, the sampling ranges for both the mechanical and coupled dataset are summarized.
They are chosen to capture all load states encountered during the macroscale simulations.
The sampling results in $N_\text{mech}=28$ and $N_\text{coup}=64$ load paths, respectively, with $M=20$ load steps each.
The invariants of the sampled load states are visualized in Figures~\ref{fig:sampled_invariants_mech} and \ref{fig:sampled_invariants_magmech}.
The invariants $\bar{I}_1^\text{iso}$ and $\bar{I}_2^\text{iso}$ as well as $\bar{I}_7^\text{iso}$ and $\bar{I}_8^\text{iso}$ are closely related, but they are not identical.

To create an RVE which is approximately transversely isotropic with the preferred direction $\bar{S}$ in $z$-direction, the central points of all inclusions are distributed on the $x$-$y$-plane, and periodic boundary conditions from Equation~\eqref{eq:periodic_bc} are applied.
A section of the RVE can be seen in Figure~\ref{fig:RVE_3D_sw} and a cross-sectional view is shown in Figure~\ref{fig:microscale_fields}.
The RVE with dimensions of $\SI{28}{\micro\meter} \times \SI{28}{\micro\meter} \times \SI{1}{\micro\meter}$ in $x$-, $y$- and $z$-direction, respectively, includes 400 particles with a diameter of $d = \SI{0.909}{\micro\meter}$, which results in a particle volume fraction of $\phi = 20\%$.\footnote{The selected length unit of \si{\micro\meter} serves illustrative purposes only, as solely the length fractions are relevant.} For the placement of the inclusions, we use the algorithm from Melro~et~al. \cite{Melro08} with a minimal distance of $0.4 \, d$ between the inclusions.

\subsubsection{Microscale material models}\label{sec:Microscale_material_models}

For both constituents, i.e., the polymer matrix and the particles, the free energy
\begin{equation}
    \label{eq:psi_mech_psi_mag_psi_vac}
    \Psi(\te{F},\ve{B}) 
    = \Psi^\text{iso}(\te{F}^\text{iso}) 
    + \Psi^\text{vol}(J) 
    + \Psi^\text{mag}(\te{F},\ve{B})
    + \Psi^\text{vac}(\te{F},\ve{B})
\end{equation}
is additively split in multiple components.
The vacuum part 
\begin{equation}
    \label{eq:psi_vac}
    \Psi^\text{vac} = \frac{1}{2 \mu_0 J} \left| \te{F} \cdot \ve{B} \right|^2 ,
\end{equation}
is independent of the material, and both components use a volumetric contribution of the form
\begin{equation}
    \label{eq:psi_vol}
    \Psi^\text{vol} = 
    \frac{\kappa}{4}\left(J^2 - 2 \op{ln}(J) - 1\right) ,
\end{equation}
where $\kappa$ is the respective bulk modulus.

For the isochoric component of the matrix, we choose the Ogden model \cite{Ogden97} with
\begin{equation}
    \label{eq:psi_ogden_iso}
    \Psi^\text{iso,M} = 
    \sum_{p=1}^3 
    \left(
    \frac{\mu_p}{\alpha_p}
    \left( (\lambda_1^\text{iso})^{\alpha_p} + 
    (\lambda_2^\text{iso})^{\alpha_p} + 
    (\lambda_3^\text{iso})^{\alpha_p} - 3 \right)
    \right) ,
\end{equation}
which is parameterized for silicone rubber with a silicone oil ratio of 20\% by Kalina~et~al. \cite{Kalina21Diss}.
The material parameters are given in Table~\ref{tab:material_params_matrix}.
As the matrix behaves like a vacuum from a magnetic perspective, the magnetic contribution $\Psi^\text{mag}(\te{F},\ve{B}) = 0$ vanishes.

\begin{table*}[]
    \begin{center}
        \small
        \caption{Material parameters of the matrix, taken from Kalina~et~al. \cite{Kalina21Diss} for a silicone oil ratio of 20\% and with an increased Poisson's ratio.}
        \begin{tabular}{ c c c c c c c c }
        \hline
            $\mu_1 / \si{\kilo\pascal}$ & $\mu_2 / \si{\kilo\pascal}$ & $\mu_3 / \si{\kilo\pascal}$ & $\alpha_1$ & $\alpha_2$ & $\alpha_3$ & $G$ / \si{\kilo\pascal} & $\nu$ \\ 
            \hline
            -11.80 & 12.45 & $4.59\cdot10^{-5}$ & -6.68 & 2.09 & 18.34 & 52.40 & 0.495 \\
            \hline
        \end{tabular}
    \label{tab:material_params_matrix}
    \end{center}
\end{table*}

To model the isochoric part of the particles, the Neo-Hooke model \cite{Ciarlet88}
\begin{equation}
    \label{eq:psi_NeoHooke_iso}
    \Psi^\text{iso,P} = 
    \frac{G}{2} \left(\op{tr}{\te{C}^\text{iso}} - 3 \right)
\end{equation}
with the material parameters $G = \SI{251.1}{\mega\pascal} $, $\nu = 0.3$, $\chi = 0.9$ and $m_\textsc{S} = \SI{1000}{\kilo\ampere\per\meter}$ is selected.
A value much smaller than the actual value of \SI{80}{\giga\pascal} is chosen for the shear modulus $G$ to increase the numerical stability. 
The chosen value is large enough to ensure that the particles are rigid compared to the matrix.
The magnetic behavior is modeled using the Langevin-type model \cite{Danas17}
\begin{equation}
    \label{eq:psi_langevin}
    \Psi^\text{mag,P}(\te{F},\ve{B}) =
    \frac{\mu_0 m_\text{s}^2}{3 \chi} 
    \left[
        \ln\left(\frac{3 \chi}{\mu_0 m_\text{s}} J^{-\frac{1}{2}} \left|\te{F}\cdot\ve{B}\right| \right)
    - \ln\left(\sinh\left(\frac{3 \chi}{\mu_0 m_\text{s}} J^{-\frac{1}{2}} \left|\te{F}\cdot\ve{B}\right| \right)\right)
    \right] ,
\end{equation}
which exhibits the saturation behavior described in Equation~\eqref{eq:magnetic_saturation}.

\subsubsection{Microscale Finite Element simulations}\label{sec:Microscale_Finite_Element_simulations}

\begin{figure*}[]
    \centering
    \subfloat[]{
        \includegraphics[width=0.32\textwidth]{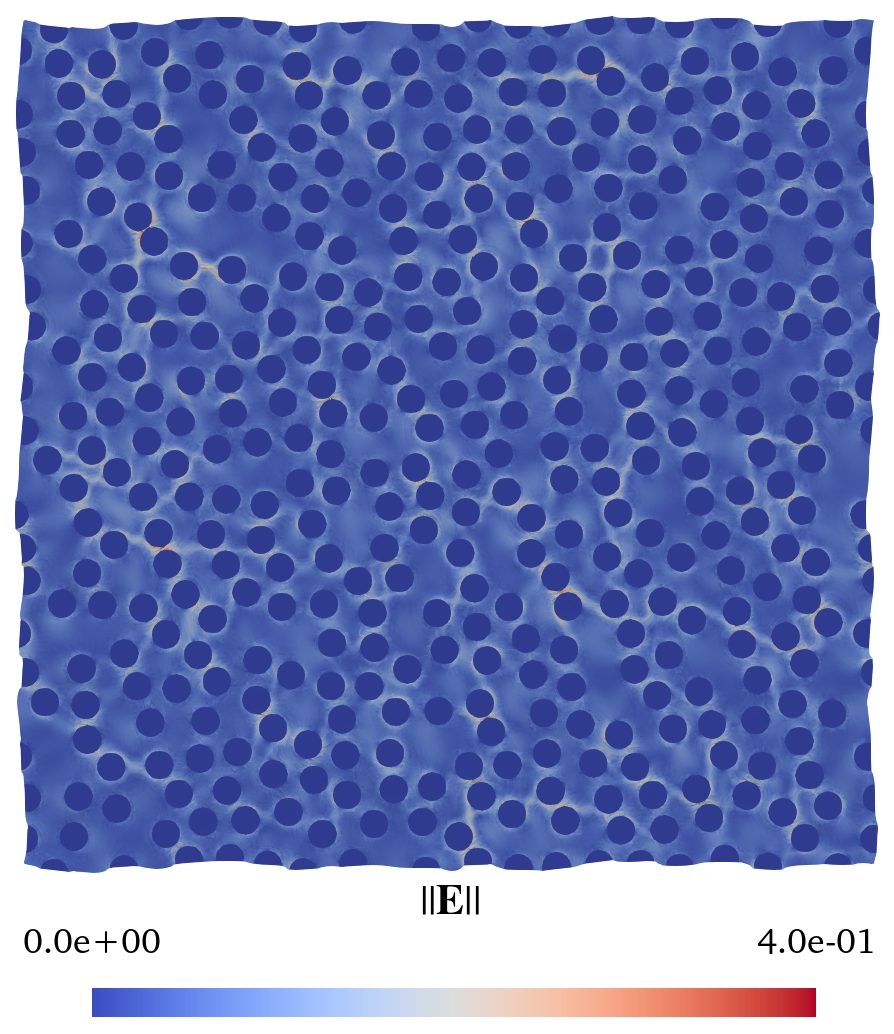}
    }
    \subfloat[]{
        \includegraphics[width=0.32\textwidth]{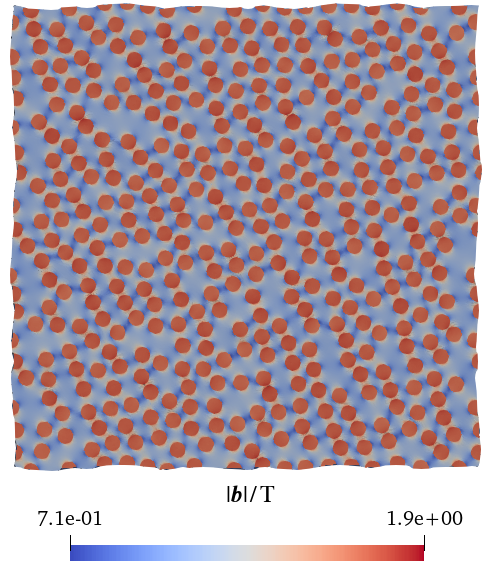}
    }
    \subfloat[]{
        \includegraphics[width=0.32\textwidth]{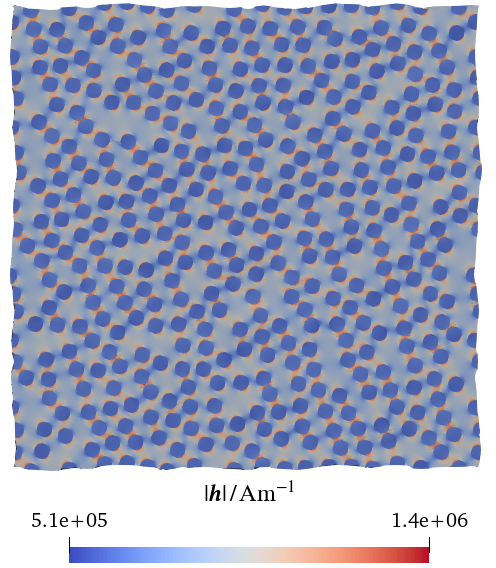}
    }
    \caption{Microscale (a) Frobenius norm of the Green-Lagrange strain tensor, (b) magnitude of the magnetic induction and (c) magnitude of the magnetic field in the $x$-$y$-plane of the RVE for one exemplary load case, obtained by an FE simulation. The RVE is distorted by the displacement $\ve{u}$ multiplied by a factor of 5.}
    \label{fig:microscale_fields}
\end{figure*}

The RVE created according to Section~\ref{sec:Load_sampling} is exposed to the load states $\{{}^a \bar{\te{F}}, {}^a \bar{\ve{B}}\}$ from Section~\ref{sec:Load_sampling}. 
By using the FE formulation from Section~\ref{sec:Variational_formulation} and the homogenization scheme described in Section~\ref{sec:Scale_transition_scheme}, we determine ${}^a \bar{\te{P}}^\text{tot}$ and ${}^a \bar{\ve{H}}$.
All meshing and FE simulations in this work are performed with the Python package \textit{Netgen/NGSolve}.\footnote{The finite element software \textit{NGSolve} and its mesher \textit{Netgen} are available at \url{https://ngsolve.org/}.}
We choose a polynomial degree of 2,2,0 and 0 for the primary field variables $\tilde{\ve{u}}$, $\tilde{\ve{A}}$, $\theta$ and $p$, respectively, with the element types given in Appendix~\ref{app:Finite_Element_discretization}.
For some loads, no convergence could be achieved during the FE simulations.
These load states were removed from the dataset, and 467 and 1125 states remain for the purely mechanical and coupled magneto-mechanical case, respectively.
The states which are included in the dataset are highlighted in Figures~\ref{fig:sampled_invariants_mech} and \ref{fig:sampled_invariants_magmech}.
For one exemplary load case, microscale field distributions are shown in Figure~\ref{fig:microscale_fields}.
The Frobenius norm of the Green-Lagrange strain tensor $\frob{\te{E}}$ within the particles is very small, as the particles have a very high stiffness.
The magnitude of the magnetic induction and magnetic field are not constant within the inclusions, especially for closely spaced particles, where both increase on adjacent sides.
A similar pattern is observed for the magnetization. 
This behavior is expected, as only a homogeneous magnetic field induces a uniform magnetization within an ellipsoidal body \cite{Takahashi}. 
In proximity to other particles, however, the homogeneity of the magnetic field is disturbed.
Furthermore, it can be seen that the particles arrange themselves in chains in the direction of the macroscale magnetic induction.
These chains are not to be confused with the chains in the preferred direction, which are created during the manufacturing process.


\subsection{Training of the PANN model}\label{sec:Training_of_the_neural_network-based_model}

\begin{figure*}[]
    \centering
    \subfloat[Losses during the training of the coupled part from 20 training runs. The best run, i.e., the run with the lowest final $\mathcal{L}^\text{cmv}_\text{train}$, is highlighted.\label{fig:losses_all}]{
        \includegraphics[width=0.45\textwidth]{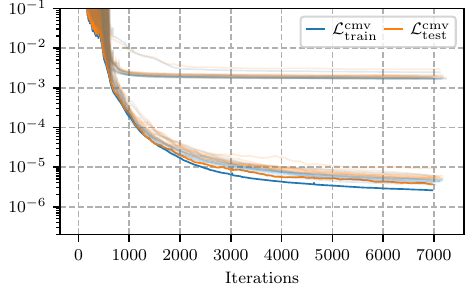}
    }
    \quad\quad
    \subfloat[Partial losses, including $\mathcal{L}^\text{smv} = \mathcal{L}^\text{sat} + \mathcal{L}^\text{mon} + \mathcal{L}^\text{con}$, for the run with the lowest final $\mathcal{L}^\text{cmv}_\text{train}$, calculated for the training and test data, respectively.\label{fig:losses_best}]{
        \includegraphics[width=0.45\textwidth]{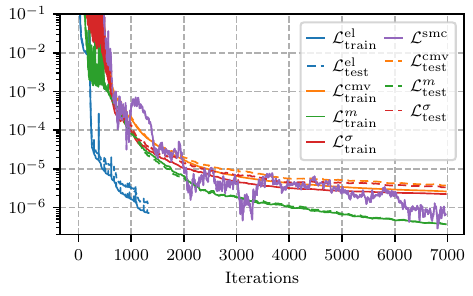}
    }
    \caption{Losses during the training of the PANN model, which consists of a pre-training with Adam followed by a post-training with SLSQP.}
    \label{fig:losses}
\end{figure*}

Utilizing the homogenization data obtained in Section~\ref{sec:Data_generation}, the training of the macroscale model is carried out according to Section~\ref{sec:Training_procedure}.
The ratio of numbers of data tuples in the training dataset to those in the test dataset is $70 / 30$.
The elastic PNN has one hidden layer with 6 neurons, and the coupled PNN has two hidden layers with 10 neurons each, resulting in a total of 276 parameters. 
In all neurons, the Softplus activation function is used.
The training of both the elastic and coupled neural network consists of a pre-training using the optimizer Adam, which is based on the stochastic gradient descent method, followed by a post-training using the quasi-newton optimizer SLSQP.
The weighting factors $w^{\te{\upsigma}}= w^{\ve{m}} = 1$ and $w^\text{sat} = w^\text{mon} = w^\text{con} = 10^{-2}$ are chosen.
For the additional dataset $\mathcal{D}^\text{add}$, the angles $^i \phi^\text{add} \in \left[0, \frac{\pi}{2}\right]$ are uniformly drawn with $n^\text{add}_\phi = 6$. 
In each direction, values $^j b^\text{add} \in \left[ \SI{2}{\tesla}, \SI{4}{\tesla} \right]$ are sampled with constant step size and  $n^\text{add}_b = 9$, 
resulting in a total of 54 additional load states.

Using this configuration, the model is trained 20 times with randomly initialized trainable variables, as shown in Figure~\ref{fig:losses_all}.
11 of these training runs reach a relatively small loss and the preferred direction $\bar{\ve{S}}$ is very accurately determined.
In the remaining 9 cases, the loss stays high, and the prediction is nearly perpendicular to the true preferred direction.
Notably, for all successful runs, the angle between the initial guess and the true preferred direction is less than 52\textdegree.
This behavior may be attributed to the directional elasticity of the material, defined in Equation~\eqref{eq:E_for_elastic_surface}: around an angle of 52\textdegree, the stiffness reaches a local minimum, while it increases monotonically toward local maxima at 0\textdegree{} and 90\textdegree, as shown in Figure~\ref{fig:elastic_magnetic_surface_slice}.
The losses do not decrease significantly with more increments, which might indicate, that a local minimum was reached.
For all runs, the loss $\mathcal{L}^{\upsigma^\text{el}}_\text{train}$ of the elastic PNN is smaller than the loss $\mathcal{L}^{\upsigma,m}_\text{train}$ from the coupled PNN, which is dominated by the term $\mathcal{L}^{\upsigma}_\text{train}$.
In the remainder of this work, the model for which the smallest final loss $\mathcal{L}^\text{cmv}_\text{train}$ is reached will be considered. 
Its partial losses are shown in Figure~\ref{fig:losses_best}, both for the training and test dataset.

\subsection{Interpolation behavior of the PANN model}\label{sec:Interpolation_behavior_of_the_neural_network-based_model}

\begin{figure}[t]
    \centerline{\includegraphics[width=0.3\columnwidth]{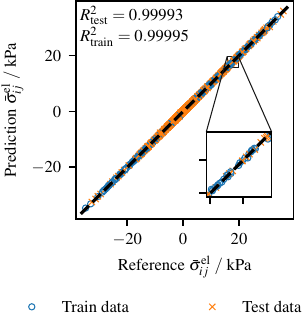}}
    \caption{Comparison of the prediction of the elastic part of the PANN model after the first training step with the reference data from the purely mechanical dataset.}
    \label{fig:correlation_el}
\end{figure}

\begin{figure*}[t]
    \centerline{\includegraphics[width=\textwidth]{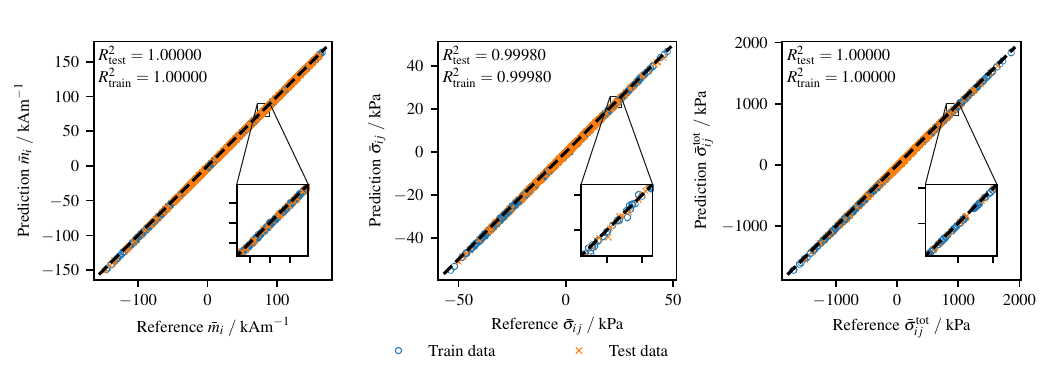}}
    \caption{Comparison of the prediction of the PANN model after the second training step with the reference data from the coupled dataset for both the training data used for model training and the test data not seen during training.}
    \label{fig:correlation_cmv}
\end{figure*}

To quantify the prediction quality of the model, the coefficient of determination is introduced as 
\begin{equation}
    R^2 = 1 - \frac{\sum_{b=1}^{n_b} \left({}^b y^\text{ref} - {}^b y^\text{pred}\right)^2}{\sum_{b=1}^{n_b} \left({}^b y^\text{ref} - y^\text{ref}_\text{avg}\right)^2} \in \mathbb{R}_{\le 1} ,
\end{equation}
where $y^\text{ref}_\text{avg} = \frac{1}{n_b} \sum_{b=1}^{n_b} {}^b y^\text{ref}$ is the average of the reference values $y^\text{ref}$, and $y^\text{pred}$ are the predictions of the model.
Tensors of higher order can be compared by arranging their coordinates in a vector.
A perfect correlation is given by $R^2 = 1$.
After the first training step with purely mechanical data, the correlation between the true value and the prediction of the coordinates $\upsigma_{ij}^\text{el}$ is plotted in Figure~\ref{fig:correlation_el} for both the train and test data. 
For both datasets, a very good correlation was achieved with $R_\text{test}^2 = 0.99993$ and $R_\text{train}^2 = 0.99995$.
Following the second training step using the coupled dataset, the accuracy of the model's magneto-mechanical predictions is evaluated in Figure~\ref{fig:correlation_cmv}.
As is to be expected from the low losses shown in Figure~\ref{fig:losses_best}, both magnetization and stress can be accurately modeled.
The coefficient of determination is excellent for the coordinates of the magnetization $\bar{m}_i$ and total stress $\bar{\upsigma}^\text{tot}_{ij}$ with $R^2 = 1.00000$ for both training and test data, respectively, and very good for the coordinates of the stress $\bar{\upsigma}_{ij}$ with $R_\text{test}^2 = 0.99980$ and $R_\text{train}^2 = 0.99980$.


\begin{figure}[]
    \centering
    \subfloat[Elastic surface\label{fig:el_surface}]{
        \includegraphics[width=0.21\columnwidth]{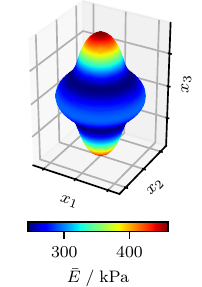}
    }
    \quad
    \subfloat[Magnetic surface\label{fig:mag_surface}]{
        \includegraphics[width=0.223\columnwidth]{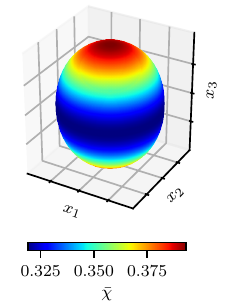}
    }
    \quad
    \subfloat[Elasticity and susceptibility depending on the polar angle\label{fig:elastic_magnetic_surface_slice}]{
        \includegraphics[width=0.45\columnwidth]{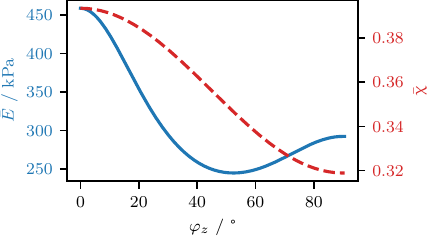}
    }
    \caption{Anisotropy of the PANN model, calculated using Equations~\eqref{eq:E_for_elastic_surface} and \eqref{eq:chi_for_magnetic_surface}, for the state $\bar{\te{F}} = \te{1}$ and $\bar{\ve{B}}= \ve{0}$ and the preferred direction $\bar{\ve{S}}=\ve{e}_3$: (a) Elastic surface, (b) magnetic surface, and (c) scalar elasticity and susceptibility depending on the angle $\phi_z$ between $\ve{n}$ and the preferred direction $\bar{\ve{S}}$.}
    \label{fig:el_mag_surface}
\end{figure}

Elastic surfaces \cite{Nordmann18} allow to visualize anisotropic elastic properties by assigning the scalar elasticity measure
\begin{equation}
    \label{eq:E_for_elastic_surface}
    \bar{E} = \left((\ve{N} \otimes \ve{n}) : \left(\frac{\partial^2 \bar{\Psi}}{\partial \bar{\te{F}} \partial \bar{\te{F}}}\right)^{\!-1} : (\ve{N} \otimes \ve{n})\right)^{\!-1}
\end{equation}
for the unloaded state to each direction vector $\ve{n} \in \mathcal{N}$.
Likewise, a scalar value for the susceptibility
can be obtained as 
\begin{equation}
    \label{eq:chi_for_magnetic_surface}
    \bar{\chi} = \left( \te{1} - \mu_0 \bar{J} \bar{\te{F}}^\text{-T} \cdot \frac{\partial^2 \bar{\Psi}}{\partial \bar{\ve{B}} \partial \bar{\ve{B}}} \cdot \bar{\te{F}}^\text{-T} \right) : (\ve{n} \otimes \ve{n}) .
\end{equation}
The elastic surface of the trained model is illustrated for the unloaded case in Figure~\ref{fig:el_surface}.
As the model is transversely isotropic, the stiffness depends only on the angle $\varphi_z = \angle(\bar{\ve{S}}, \ve{n})$ to the preferred direction, with a minimal and maximal stiffness being achieved for $\varphi_z=\ang{0}$ and approximately 52\textdegree, respectively, as shown in Figure~\ref{fig:elastic_magnetic_surface_slice}. 
Compared to the initial Young's modulus of the matrix of \SI{156.68}{\kilo\pascal}, the composite material is significantly stiffer with $\bar{E} \in \left[ \SI{244}{\kilo\pascal}, \SI{459}{\kilo\pascal} \right].$
The magnetic anisotropy in Figure~\ref{fig:mag_surface} is far less pronounced, but there is still a noticeable increase in the magnetic susceptibility, when the magnetic far field is more aligned with the direction of the particle chains.

\begin{figure}[t]
\centering
\includegraphics[width=0.45\columnwidth]{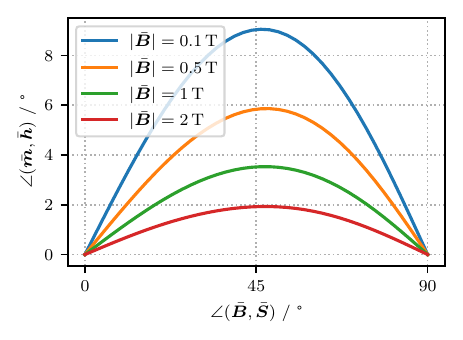}
\caption{Angle between $\bar{\ve{m}}$ and $\bar{\ve{h}}$ predicted by the PANN model for $\bar{\te{F}}=\te{1}$, depending on the magnitude of the magnetic induction and the angle between $\bar{\ve{B}}$ and $\bar{\ve{S}}$. The magnetization deviates from the magnetic induction in the preferred direction, which leads to $\angle(\bar{\ve{m}},\bar{\ve{S}}) \le \angle(\bar{\ve{b}},\bar{\ve{S}}) \le \angle(\bar{\ve{h}},\bar{\ve{S}})$.}
\label{fig:angle_m_h}
\end{figure}

In isotropic materials, the vectors $\bar{\ve{b}}$, $\bar{\ve{h}}$ and $ \bar{\ve{m}}$ are always aligned in the undeformed state.
In contrast, for the structured MRE studied here, this holds only when $\angle(\bar{\ve{S}},\bar{\ve{B}}) \in \{\ang{0}, \ang{90} \}$.
At other orientations, the directions of $\bar{\ve{m}}$ and $\bar{\ve{h}}$ deviate from one another, with the angle between them decreasing as the magnetic induction increases, as demonstrated in Figure~\ref{fig:angle_m_h}.

\subsection{Extrapolation behavior of the PANN model}\label{sec:Extrapolation_behavior_of_the_neural_network-based_model}

\begin{figure}[t]
\centering
\includegraphics[width=0.5\columnwidth]{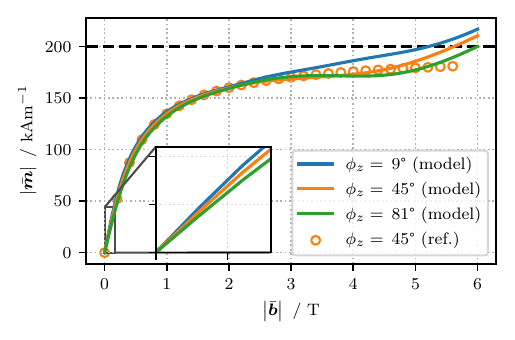}
\caption{Extrapolation of the magnetic saturation curve for $\bar{\ve{b}}$ in different directions, specified by the angle $\phi_z$ between the preferred direction $\bar{\ve{S}}$ and magnetic induction $\bar{\ve{B}}$ with $\te{F}=\te{1}$. The saturation magnetization is $m_\text{s} = \SI{200}{\kilo\ampere\per\meter}$. Magnetic inductions from 0 to \SI{2.2}{\tesla} are included within the training dataset. An additional dataset, which enforces conditions on the magnetization curve in a weak sense, covers $\SI{2}{\tesla} \le \norm{\bar{\ve{b}}} \le \SI{4}{\tesla}$. For one angle, reference data from an RVE simulation is shown.}
\label{fig:magnetization_extrapolation}
\end{figure}

Several conditions on the magnetic saturation curve are considered by enforcing boundedness, monotony, and concavity of an extended dataset during the training.
To evaluate the success of this approach, the magnetization curves in three directions, which were not included in the additional training dataset, are determined in Figure~\ref{fig:magnetization_extrapolation}.
Within the range of the coupled dataset, i.e., for $\norm{\bar{\ve{B}}} < \SI{2.2}{T}$, the conditions mentioned above are met without requiring an additional loss term.
As suggested by Figure~\ref{fig:elastic_magnetic_surface_slice}, the initial susceptibility, which is proportional to the initial slope of the magnetization curve, is higher for lower angles between $\bar{\ve{S}}$ and $\bar{\ve{B}}$.
Within the additional data range, i.e., for $\norm{\bar{\ve{B}}} \in \left[ \SI{2}{\tesla}, \SI{4}{\tesla} \right]$, the three magnetization curves begin to diverge more noticeably, although they still approximately satisfy the imposed conditions. At higher magnetic inductions, however, both the magnetic saturation and concavity conditions are substantially violated.

The volumetric growth condition, explicitly incorporated into the model, has been numerically verified for $\bar{J} \to 0.01$ and $\bar{J} \to 100$.
Although not enforced explicitly, the magnetic growth condition is also satisfied in a numerical test with $\norm{\bar{\ve{B}}} \to \SI{20}{\tesla}$.
To numerically validate the positivity of the free energy, $10^5$ random load states in the ranges $\bar{\lambda}_i \in [0.2, 5]$, $\bar{J} \in [0.9, 1.1]$ and $\norm{\bar{\ve{B}}} \in [\SI{0}{\tesla}, \SI{10}{\tesla}]$ are generated.
Among these, $0.025\%$ result in a negative free energy, with the magnitude of the negative values being significantly lower than the majority of the positive values.

\subsection{Macroscale magnetostrictive effect}\label{sec:Macroscale_magnetostrictive_effect}

\begin{figure*}[]
    \centering
    \subfloat[$\bar{\ve{S}} \parallel \ve{e}_3$]{
        \includegraphics[width=0.4\textwidth]{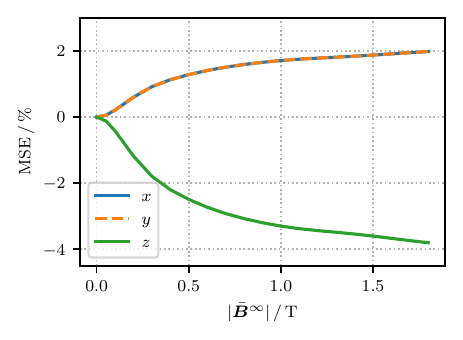}
    }
    \quad\quad
    \subfloat[$\bar{\ve{S}} \parallel \ve{e}_1$]{
        \includegraphics[width=0.4\textwidth]{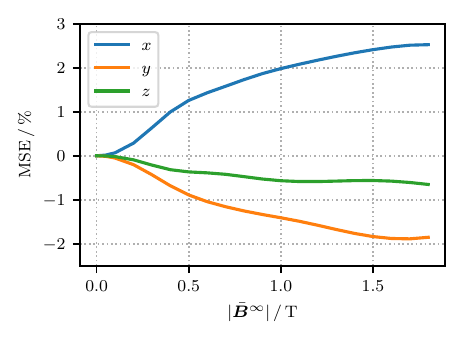}
    }
    \caption{Magnetostrictive effect with (a) the preferred direction $\bar{\ve{S}}$ parallel and (b) perpendicular to the homogeneous magnetic far field $\bar{\ve{B}}^\infty = \ve{e}_3$, respectively.}
    \label{fig:MSE}
\end{figure*}

\begin{figure*}[]
    \centering
    \subfloat[A spherical macroscopic MRE sample is embedded in a cubical vacuum domain and exposed to a homogeneous magnetic far field. $\bar{\ve{B}}^\infty = (1.5 \ve{e}_3) \si{\tesla}$.\label{fig:MSE_z_z_1.5T_b_3D}]{
        \includegraphics[width=0.45\textwidth]{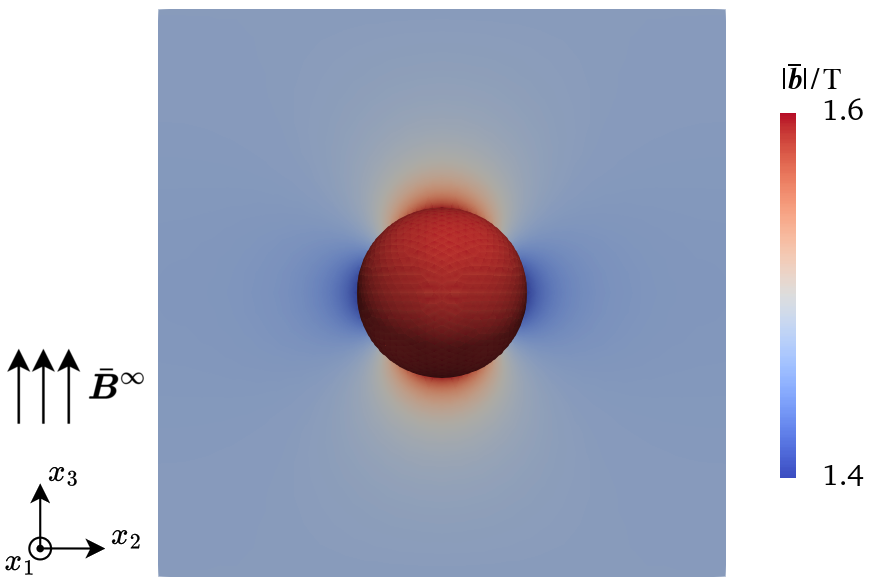}
    }
    \quad\quad
    \subfloat[Magnetostriction of a spherical macroscopic MRE sample caused by the homogeneous magnetic far field $\bar{\ve{B}}^\infty = (1 \ve{e}_3) \si{\tesla}$. A black line marks the original geometry. The model is validated for load states at points A, B and C.\label{fig:MSE_with_points}]{
        \includegraphics[width=0.4\textwidth]{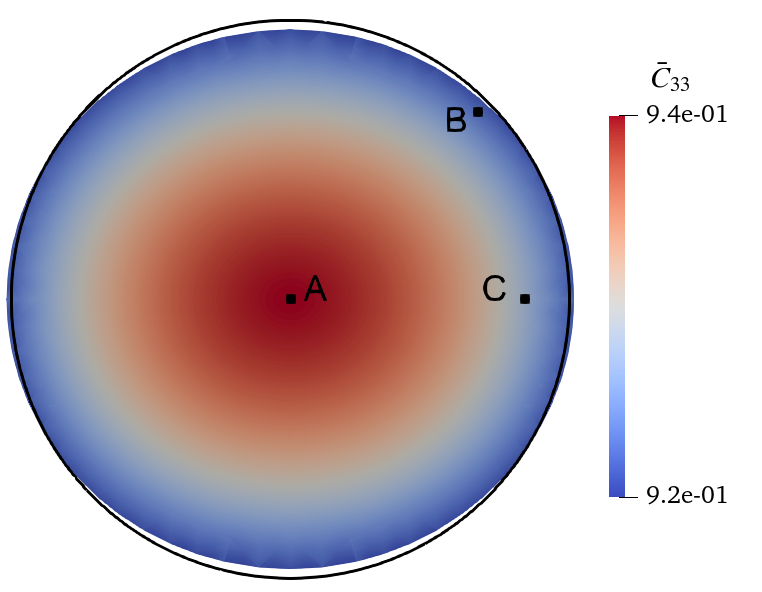}
    }
    \caption{A homogeneous magnetic far field is applied parallel to the chain direction $\bar{\ve{S}} = \ve{e}_3$ of a spherical macroscopic MRE sample to quantify the magnetostrictive effect.}
\end{figure*}

\begin{figure*}[]
    \centering
    \subfloat[Point A]{
        \includegraphics[width=0.32\textwidth]{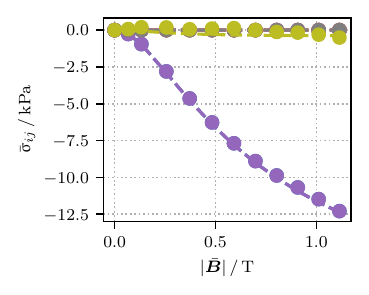}
    }
    \subfloat[Point B]{
        \includegraphics[width=0.32\textwidth]{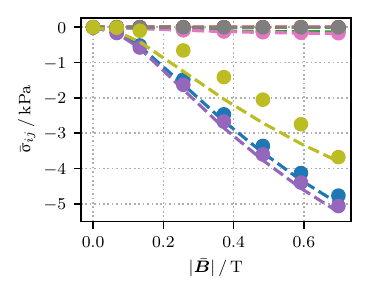}
    }
    \subfloat[Point C]{
        \includegraphics[width=0.32\textwidth]{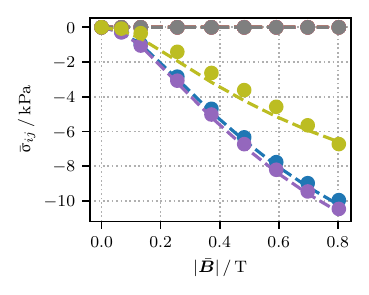}
    }
    \vspace{0.5em} 
    \includegraphics[width=0.9\textwidth]{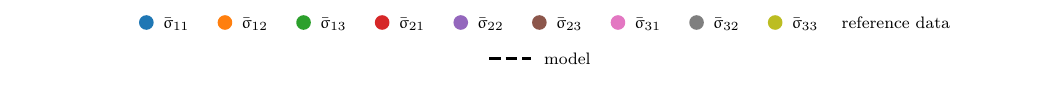}
    \caption{Comparison of the stress prediction of the PANN model at points (a) A, (b) B and (c) C (see Figure~\ref{fig:MSE_with_points}) with reference data from RVE simulations. The range of the magnetic induction is restricted due to convergence failures in the microscale RVE simulations.}
    \label{fig:localization_MSE}
\end{figure*}

A spherical macroscopic MRE sample $\bar{\mathcal{B}}^\text{mat}_0 = \left\{ \ve{X} \in \mathbb{R}^3 \ | \ \norm{\ve{X}} \le r \right\}$ with a radius of $r = \SI{3}{\centi\meter}$, whose behavior is described by the PANN model obtained in Section~\ref{sec:Training_of_the_neural_network-based_model}, is embedded in a cubical pseudo vacuum domain $\bar{\mathcal{B}}_0^\text{vac} = \left\{ \ve{X} \in \mathbb{R}^3 \ | \ \norm{X_i} \le \frac{l^\text{vac}}{2} \right\} \setminus \bar{\mathcal{B}}^\text{mat}_0 $ with a length of $l^\text{vac} = \SI{20}{\centi\meter}$ and subjected to a homogeneous magnetic far field $\bar{\ve{B}}^\infty$, as shown in Figure~\ref{fig:MSE_z_z_1.5T_b_3D}.
The free energy of the pseudo vacuum is given by 
$\bar{\Psi}^\text{PV}(\bar{\te{F}}, \bar{\ve{B}}) = \bar{\Psi}^\text{PV,mech}(\bar{\te{F}}) + \bar{\Psi}^\text{vac}(\bar{\te{F}}, \bar{\ve{B}})$, where
\begin{equation}
    \bar{\Psi}^\text{PV,mech}(\bar{\te{F}}) = \frac{G_\text{vac}}{2} \left( \lVert {\bar{\te{F}}^\text{iso}} \rVert ^2 - 3 \right) + \frac{\kappa_\text{vac}}{4} \left(\bar{J}^2 - 2\ln \bar{J} - 1 \right)
\end{equation}
with $G_\text{vac} = \SI{200}{\pascal}$ and $\kappa_\text{vac} = \SI{200}{\pascal}$.
The magnetostrictive effect describes a change of length caused by a magnetic field and is defined as $\mathrm{MSE} = \frac{\Delta l}{l_0}$, where $l_0$ and $\Delta l$ are the initial length and change of length in the respective direction, respectively.
By utilizing the FE formulation described in Section~\ref{sec:Macroscale_simulation} and choosing a polynomial degree of 3, 3, 2 and 2 for the primary field variables $\bar{\ve{u}}$, $\bar{\ve{A}}$, $\theta$ and $p$, respectively, the magnetostrictive effect is obtained for both $\bar{\ve{S}} = \ve{e}_3$ and $\bar{\ve{S}} = \ve{e}_1$, while $\bar{\ve{B}}^\infty \parallel \ve{e}_3$, as shown in Figure~\ref{fig:MSE}.

For $\bar{\ve{S}} = \ve{e}_3$, i.e., $\bar{\ve{S}} \parallel \bar{\ve{B}}^\infty$, the sample contracts in field direction and expands perpendicular to it.
Simulations and experiments for specific particle arrangements show the same qualitative behavior \cite{kalina_microscale_2016,Danas17,Fischer24,Fischer24_2,Martin06}. 
This is in contrast to isotropic MREs, where the opposite effect is reported \cite{Guan07,Kalina21Diss,Gebhart24Diss}.
At $\norm{\bar{\ve{B}}^\infty} = \SI{1.8}{\tesla}$, the magnetostrictive effect parallel and perpendicular to the direction of the far field is approximately $-4\%$ and $2\%$, respectively.
The resulting load states with $0.953 < \bar{\lambda}_i < 1.027$, $0.9986 < \bar{J} < 1.0014$, and $|\bar{\ve{B}}| < \SI{2.119}{\tesla}$ lie within the sampling ranges noted in Table \ref{tab:sampling_params}, though the minimum principal stretches approach the lower boundary of $0.95$.
Macroscale fields in the $y$-$z$-plane are shown in Appendix~\ref{app:Field_distributions_during_magnetostriction}.

For $\bar{\ve{S}} = \ve{e}_1$, i.e., $\bar{\ve{S}} \perp \bar{\ve{B}}^\infty$, the simulation predicts an expansion of up to $2.5\%$ in $x$-direction, a contraction of up to $-2\%$ in $y$ direction and slight contraction in $z$-direction.
Although the resulting load states $0.964 < \bar{\lambda}_i < 1.037$, $0.9992 < \bar{J} < 1.0021$, and $|\bar{\ve{B}}| < \SI{2.019}{\tesla}$ fall within the sampling ranges, the maximum principal stresses come close to the maximum sampling threshold of $1.04$.

To further validate the model, the states $\bar{\ve{F}}$ and $\bar{\ve{B}}$ are stored at the three integration points: A $(0,0,0)\,\si{\centi\meter}$, B $(2,0,2)\,\si{\centi\meter}$ and C $(2.5,0,0)\,\si{\centi\meter}$, which are shown in Figure~\ref{fig:MSE_with_points}. 
The used coordinate system originates in the center of the sphere.
Using microscale RVE simulations as outlined in Section~\ref{sec:Microscale_Finite_Element_simulations}, the corresponding mechanical stresses are obtained to be used as reference values in the following.
As shown in Figure~\ref{fig:localization_MSE}, in point A, the model accurately predicts the stress. 
In point B, $\bar{\upsigma}_{33}$ deviates from the reference values.
This might be caused by the small magnitude of the stress coordinates at point B, which are less than \SI{5}{\kilo\pascal}, compared to the training data, which includes mechanical stresses up to \SI{50}{\kilo\pascal}.
In point C, $\bar{\upsigma}_{11}$ and $\bar{\upsigma}_{22}$ are accurately predicted, while there are minor deviations in $\bar{\upsigma}_{33}$.
In summary, the proposed model reliably captures stresses of sufficiently large magnitude, supporting the validity of the obtained magnetostriction behavior.

\section{Conclusion}\label{chap:Conclusion}

In the present work, a data-driven multiscale scheme for structured, elastic, and magnetically soft MREs is developed.
A macroscale PANN model is trained using homogenization data from microscale RVE simulations, validated, and deployed in a macroscale FE simulation to obtain the magnetostrictive effect of a sample.

To begin with, finite strain kinematics and the Maxwell equations and balance equations are summarized and the principles of constitutive modeling are given.
This is followed by a homogenization scheme and the introduction of a 4-field variational formulation, which serves as the basis of the FE simulations.

The multiscale scheme starts with the sampling of magneto-mechanical load states, which are required for the training of the macroscale model.
The transversely isotropic PANN model utilizes two PNNs and predicts the free energy.
Several physical conditions are either directly integrated in the model in a strong sense or enforced during the training in a weak sense.
These include thermodynamic consistency, objectivity, material symmetry, symmetry of the total stress tensor, vanishing stress and free energy in the unloaded state, the absence of magnetization when no magnetic field is present, the magnetic growth condition and several conditions on the magnetization curve.
During the two-step training, the preferred direction of the material is detected.
For the macroscale simulation, we adapt the FE formulation slightly.

To show the feasibility of the approach, an RVE consisting of spherical particles with center points arranged on a plane is subjected to magneto-mechanical loads.
The resulting data is used for the training of the macroscale model, which leads to an excellent prediction of magnetization, total stress and mechanical stress, assuming a good initialization of the preferred direction.
The resulting model has a transversely isotropic mechanical and magnetic behavior, with the mechanical anisotropy being far more pronounced than the magnetic anisotropy.
A reasonable extrapolation of the magnetization curve is achieved by considering additional loss functions during the training.
The obtained model is integrated in a macroscale FE simulation, where a spherical MRE sample is placed in a homogeneous magnetic far field.
If the magnetic far field and the preferred direction of the material are aligned, a contraction of up to 4\% in field direction occurs.
Finally, selected load states from the macroscale simulation are used as input of microscale RVE simulations.
The resulting responses closely match the model’s predictions, providing further validation of the proposed approach.

Several extensions of the shown approach are planned in the future.
Firstly, more realistic RVEs with a more irregular particle arrangement are to be created using microstructure reconstruction \cite{Seibert24}, which might lead to an opposite magnetostrictive effect, i.e., expansion in field direction \cite{Danas17,kalina_microscale_2016}.
This may require the extension of the model to more complex anisotropy classes \cite{Kalina24_anisotropy}.
The relation between microstructure and macroscopic behavior can be examined, and microstructures which maximize magnetic macroscopic effects can be determined.
Furthermore, the constitutive model can be used to examine the magnetorheological effect through macroscale FE simulations.
Finally, the framework could be adapted to a dissipative material by considering magnetically hard particles and magneto-viscoelasticity \cite{Gebhart24Diss,Saxena13}.

\section*{Author contributions}

\textbf{Heinrich T. Roth}: Conceptualization, Formal analysis, Investigation, Methodology, Visualization, Software, Validation, Writing – original draft, Writing – review \& editing.
\textbf{Philipp Gebhart}: Conceptualization, Formal analysis, Methodology, Writing – review \& editing.
\textbf{Karl A. Kalina}: Conceptualization, Formal analysis, Methodology, Writing – review \& editing.
\textbf{Thomas Wallmersperger}: Resources, Writing – review \& editing.
\textbf{Markus K\"{a}stner}: Resources, Writing – review \& editing.

\section*{Acknowledgment}
All presented computations were performed on a PC-Cluster at the Center for Information Services and High Performance Computing (ZIH) at TU Dresden. The authors thus thank the ZIH for generous allocations of computer time.
We thank the Deutsche Forschungsgemeinschaft (German Research
Foundation, DFG) for support  through the Research Unit FOR 5599 on
structured magnetic elastomers, project no. 511114185, via DFG grant
reference no. KA 3309/22-1.

\section*{Conflict of interest}

The authors declare no potential conflict of interests.

\appendix

\section{Function spaces and Finite Element discretization}\label{app:Finite_Element_discretization}

This section serves as supplemental material to the microscale variational formulation in Section~\ref{sec:Variational_formulation}.
In Equation~\eqref{eq:Pi_micro}, the potential $\Pi(\tilde{\ve{u}}, \tilde{\ve{A}}, \theta, p)$ is defined.
The primary variables are determined by the extremum principle
\begin{equation}
    (\tilde{\ve{u}}, \tilde{\ve{A}}, \theta, p) =
    \op{arg} \left(
        \underset{\tilde{\ve{u}} \in \mathcal{V}_{\tilde{\ve{u}}}}{\op{inf}} \ 
        \underset{\tilde{\ve{A}} \in \mathcal{V}_{\tilde{\ve{A}}}}{\op{inf}} \ 
        \underset{\theta \in \mathcal{V}_{\theta}}{\op{inf}} \ 
        \underset{p \in \mathcal{V}_{p}}{\op{sup}} \ 
        \Pi(\tilde{\ve{u}}, \tilde{\ve{A}}, \theta, p)
    \right) ,
\end{equation}
where the admissible function spaces are given as
\begin{equation}
    \begin{aligned}
        \label{eq:continuous_function_space}
        \mathcal{V}_{\tilde{\ve{u}}} &= \{ {\tilde{\ve{u}}} \in H^1(\mathcal{B}_0) \ |\ \periodplus{\tilde{\ve{u}}} = \ve{0} \text{ on } \partial \mathcal{B}_0 \} , \\
        \mathcal{V}_{\tilde{\ve{A}}} &= \{ {\tilde{\ve{A}}} \in H(\text{Curl},\mathcal{B}_0) \ |\ \ve{N} \times \periodplus{\tilde{\ve{A}}} = \ve{0} \text{ on } \partial \mathcal{B}_0 \} , \\
        \mathcal{V}_{\theta} &= \{ {\theta} \in L^2(\mathcal{B}_0) \} \text{ and}\\
        \mathcal{V}_{p} &= \{ {p} \in L^2(\mathcal{B}_0) \} .
    \end{aligned}
\end{equation}
For the definition of the function spaces above, the $L^2$ norm of a vector-valued function $\ve{a} : \mathcal{B}_0 \to \mathcal{L}_1$ and a matrix-valued function $\te{a} : \mathcal{B}_0 \to \mathcal{L}_2$ are  introduced as 
\begin{equation}
    \frob{\ve{a}}_{L^2} = \sqrt{\int_{\mathcal{B}_0} \ve{a} \cdot \ve{a} \ \dd V}
    \quad \text{and} \quad 
    \frob{\te{a}}_{L^2} = \sqrt{\int_{\mathcal{B}_0} \te{a} : \te{a}^\text{T} \ \dd V} ,
\end{equation}
respectively. Based on these norms, the Sobolev spaces
\begin{equation}
    \begin{aligned}
        L^2(\mathcal{B}_0) &= \{\ve{a} : \mathcal{B}_0 \to \mathbb{R}^3 \ |\ \frob{\ve{a}}_{L^2} < \infty\} , \\
        H^1(\mathcal{B}_0) &= \{\ve{a} \in L^2(\mathcal{B}_0) \ |\ \Grad{\ve{a}} \in L^2(\mathcal{B}_0)\} \text{ and} \\
        H(\text{Curl},\mathcal{B}_0) &= \{\ve{a} \in L^2(\mathcal{B}_0) \ |\ \op{Curl}\ve{a} \in L^2(\mathcal{B}_0)\}
    \end{aligned}
\end{equation}
describe the spaces, in which the function $\ve{a}$ and additionally its gradient and curl are square-integrable, respectively \cite{Gebhart24Diss}.
The optimality condition for the variational problem is given by the first variation $\delta\Pi = 0$, which results in the weak form of the problem \cite{Gebhart24Diss}
\begin{equation}
    \label{eq:weak_form}
        0 = \int_{\mathcal{B}_0}
        \pd{\Psi}{\te{F}} : (\Grad \delta \tilde{\ve{u}})^\text{T} +
        \pd{\Psi}{\ve{B}} \cdot (\Rot \delta\tilde{\ve{A}}) 
        + \kappa \tilde{\ve{A}} \cdot \delta\tilde{\ve{A}}
        + \left(\pd{\Psi}{\theta} - p\right) \delta \theta + 
        (J - \theta) \delta p 
        \ \dd V ,
\end{equation}
where 
$\delta \tilde{\ve{u}} \in \mathcal{V}_{\tilde{\ve{u}}}$, 
$\delta \tilde{\ve{A}} \in \mathcal{V}_{\tilde{\ve{A}}}$, 
$\delta \theta \in \mathcal{V}_{\theta}$ and
$\delta p \in \mathcal{V}_{p}$
are the variations of the primary fields.
With partial integration and the fundamental lemma of calculus applied to Equation~\eqref{eq:weak_form}, the Euler-Lagrange equations are obtained in Equation~\eqref{eq:Euler_Lagrange}.

In order to numerically solve the extremum problem, a triangulation $\mathcal{T}_h = \{T\}$ is applied to the domain $\underline{\mathcal{B}}_0 \in \mathbb{R}^3$.
The solution is now to be found in the discrete finite element spaces
\begin{equation}
    \begin{aligned}
        \mathcal{V}_{\tilde{\ve{u}}}^h &= \{ {\tilde{\ve{u}}}_h \in H^1(\underline{\mathcal{B}}_0) \ |\ \tilde{\ve{u}}_{h|_T} \in \mathcal{P}_k \ \forall T \in \mathcal{T}_h \} \subset \mathcal{V}_{\tilde{\ve{u}}} , \\
        \mathcal{V}_{\tilde{\ve{A}}}^h &= \{ {\tilde{\ve{A}}}_h \in H(\text{Curl}, \underline{\mathcal{B}}_0) \ |\ \tilde{\ve{A}}_{h|_T} \in \mathcal{N}_k^{II} \ \forall T \in \mathcal{T}_h \} \subset \mathcal{V}_{\tilde{\ve{A}}} , \\
        \mathcal{V}_{\theta}^h &= \{ {\theta}_h \in L^2(\underline{\mathcal{B}}_0) \ |\ {\theta}_{h|_T} \in \mathcal{P}_{l} \ \forall T \in \mathcal{T}_h \} \subset \mathcal{V}_{\theta} \text{ and} \\
        \mathcal{V}_{p}^h &= \{ {p}_h \in L^2(\underline{\mathcal{B}}_0) \ |\ {p}_{h|_T} \in \mathcal{P}_{l} \ \forall T \in \mathcal{T}_h \} \subset \mathcal{V}_{p} \\
    \end{aligned}
\end{equation}
for both the primary field variables and their variations \cite{Gebhart24Diss,Gebhart24}.
Here, $\mathcal{P}_k$ stands for the polynomial space of order $k$ and $\mathcal{N}_k^{II}$ for the Nédélec space \cite{Nedelec1980} of the second kind of order $k$. 
In this work, Lagrange elements of degree 2 and 3 are used for $\tilde{\ve{u}}_h$, Nédélec elements of the second kind and degree 2 and 3 are used for $\tilde{\ve{A}}_h$, and discontinuous Lagrange elements of degree 0 and 1 are used for $\theta_h$ and $p_h$ during the microscale and macroscale simulations, respectively.
These discrete spaces fulfill conformity, i.e., the discrete finite element space is included in the function space, in which the variational problem is posed \cite{Gebhart24Diss}.

\section{Data sampling}\label{app:Data_sampling}

To obtain a dataset covering a wide range of the invariants space, we deploy a sampling approach \cite{Kalina24,Kalina24_anisotropy}.
From the principles of time reversal \eqref{eq:time_reversal}, material frame invariance \eqref{eq:objectivity} and material symmetry \eqref{eq:mat_symmetry}, the condition
\begin{equation}
    \bar{\Psi}(\bar{\te{F}},\ve{\bar{B}}) = 
    \bar{\Psi}(\te{Q}_1 \cdot \bar{\te{F}} \cdot \te{Q}_2^\text{T}, \ve{\bar{B}} \cdot \te{Q}_2^\text{T})
    \ \forall \ \bar{\te{F}} \in \mathcal{GL}^+(3), \ve{\bar{B}} \in \mathcal{L}_1,
    \te{Q}_1 \in \mathcal{SO}(3), \te{Q}_2 \in \mathcal{G}^\parallel   
\end{equation}
follows.
The symmetry group of the transversely isotropic material considered here with the preferred direction $\bar{\ve{S}}$ is 
\begin{equation}
    \mathcal{G}^\parallel = \{ \te{Q} \in \mathcal{L}_2 \ |\ 
        \te{Q}(\alpha) = \cos(\alpha)\te{1} + (1 - \cos(\alpha)) \bar{\ve{S}} \otimes \bar{\ve{S}} - \sin(\alpha) \mathfrak{e} \cdot \bar{\ve{S}} \}
\end{equation}
with the permutation pseudo tensor $\mathfrak{e}$ \cite{Ebbing10}.
This means, that the material behavior is invariant with respect to rotations by an arbitrary angle $\alpha$ around $\bar{\ve{S}}$.
For sampling, the direction $\ve{\hat{\bar{S}}}{}^\text{samp} = \ve{e}_3$ and the rotation tensor $\samp{\bar{\te{R}}} = \te{1}$ are chosen, which results in $\samp{\bar{\te{F}}} = \samp{\bar{\te{U}}}$.
Different deformation states 
\begin{equation}
    \samp{\bar{\te{U}}} = \te{Q}_{\bar{\te{U}}} \cdot \op{diag}\left({\bar{\lambda}_1}, {\bar{\lambda}_2}, \frac{{\bar{J}}}{{\bar{\lambda}_1} {\bar{\lambda}_2}}\right) \cdot \te{Q}_{\bar{\te{U}}}^\text{T}
\end{equation}
are sampled with the rotation tensor $\te{Q}_{\bar{\te{U}}}(\vartheta_1, \vartheta_2)$, 
which is expressed with the Euler angles $\vartheta_1, \vartheta_2$ around two axes perpendicular to $\bar{\ve{S}}$.
Note that the sampled principal stretches $\bar{\lambda}_1$ and $\bar{\lambda}_2$ are not necessarily ordered.
The rotation of the magnetic induction of magnitude $\bar{B}$ can be expressed using rotation tensor $\te{Q}_{\bar{\ve{B}}}(\phi_1, \phi_2)$ with $\phi_1$ and $\phi_2$ as the polar and azimuthal angle with respect to $\bar{\ve{S}}$, resulting in
\begin{equation}
     \samp{\ve{\bar{B}}}  = \bar{B} \te{Q}_{\bar{\ve{B}}} \cdot \bar{\ve{S}} .
\end{equation}
To summarize, the parameters $(\bar{\lambda}_1, \bar{\lambda}_2, \bar{J}, \vartheta_1, \vartheta_2, \phi_1, \phi_2, \bar{B})$ have to be sampled within selected ranges in the magneto-mechanical case.
For the purely mechanical case, which is required as well for the training according to Section~\ref{sec:Training_procedure}, $\samp{\ve{\bar{B}}} \equiv \ve{0}$ applies, so only the parameters $(\bar{\lambda}_1, \bar{\lambda}_2, \bar{J}, \theta_1, \theta_2)$ remain.
The sampling results values for $^{n}\bar{F}_{ij}$ and ${}^n B_k $, where $n \in \mathbb{N}$ denotes the sampling index with $1 \le n \le N$ and $N$ the total number of sampled states.
For each sampled state, a loading path with increments $1 \le m \le M$ is created as 
\begin{equation}
    \label{eq:sampling_loadpaths}
    {}^{n,m}\bar{F}_{ij} = \delta_{ij} + \frac{m}{M} \left( {}^{n,M}\bar{F}_{ij} - \delta_{ij} \right)
    \quad \text{and} \quad
    {}^{n,m}\bar{B}_{k} = \frac{m}{M} {}^{n,M}\bar{B}_{k} .
\end{equation}
The last loading state is given as ${}^{n,M}\bar{F}_{ij} = {}^{n}\bar{F}_{ij}$ and ${}^{n,M}\bar{B}_{k} = {}^{n}\bar{B}_{k}$.
The total set of load states is given by $\{{}^a \bar{\te{F}}, {}^a \bar{\ve{B}}\}$ with $a \in \{1,2,\dots,M \cdot N\}$ and the corresponding invariants ${}^a \bar{I}_l$ with $l \in \{1,2,\dots,10\}$.
To reduce redundancy, the invariant based filtering described in \cite{Kalina24} is used, which removes load states, that have similar invariants, as they do not contribute to a better training.
Additionally, states with a principal stretch outside the selected sampling range for $\bar{\lambda}_i$ are excluded.

\section{Field distributions during magnetostriction}\label{app:Field_distributions_during_magnetostriction}

\begin{figure*}[]
    \centering
    \subfloat[$\bar{\upsigma}_{11} \ / \ \si{\kilo\pascal}$]{
        \includegraphics[width=0.3\textwidth]{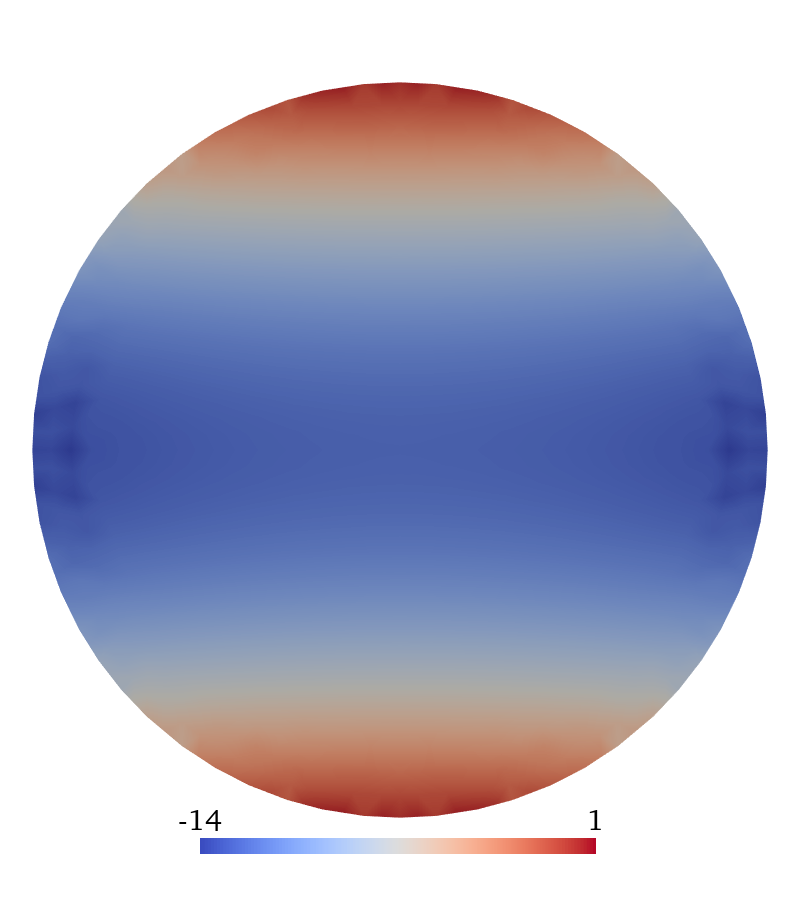}
    }
    \subfloat[$\bar{\upsigma}_{22} \ / \ \si{\kilo\pascal}$]{
        \includegraphics[width=0.3\textwidth]{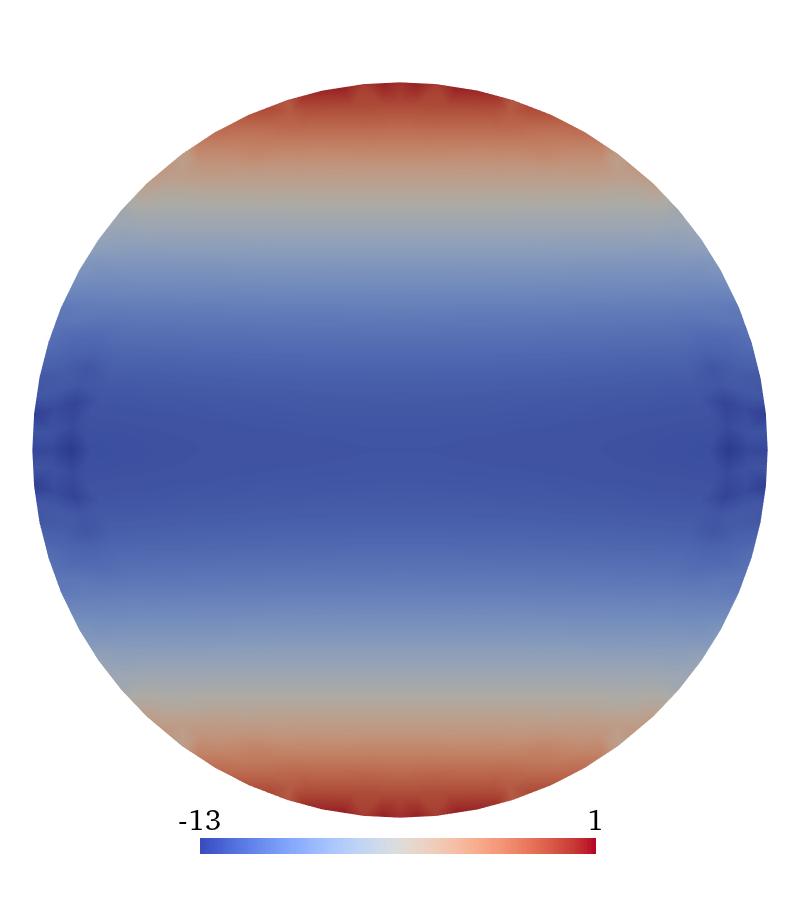}
    }
    \subfloat[$\bar{\upsigma}_{33} \ / \ \si{\kilo\pascal}$]{
        \includegraphics[width=0.3\textwidth]{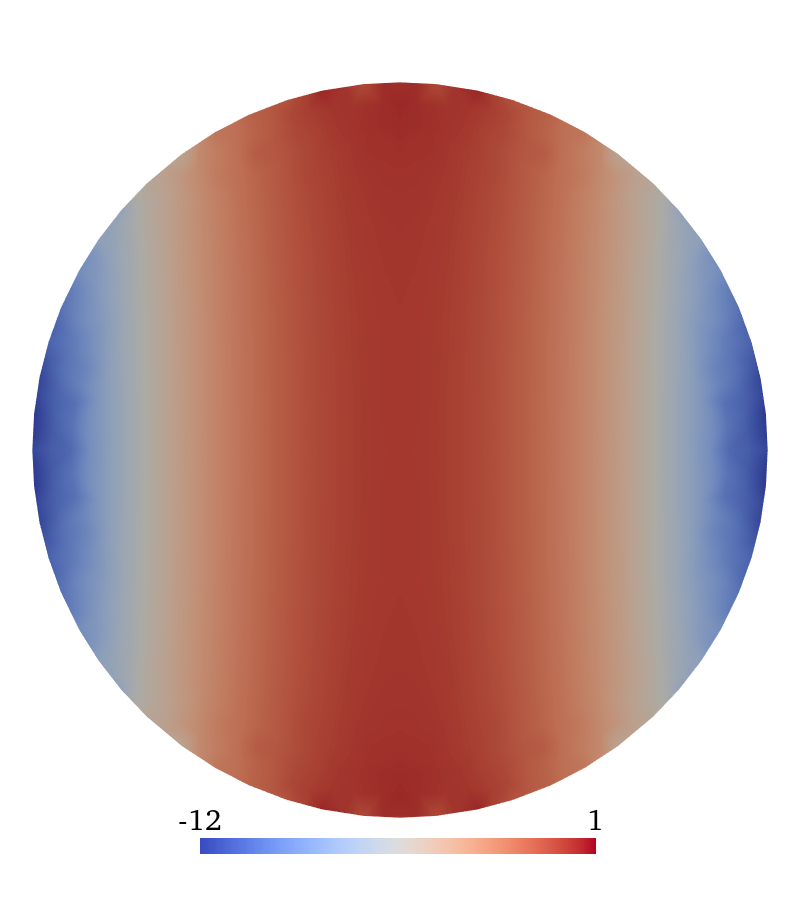}
    }
    \\
    \subfloat[$\bar{\mathrm{C}}_{11}$]{
        \includegraphics[width=0.3\textwidth]{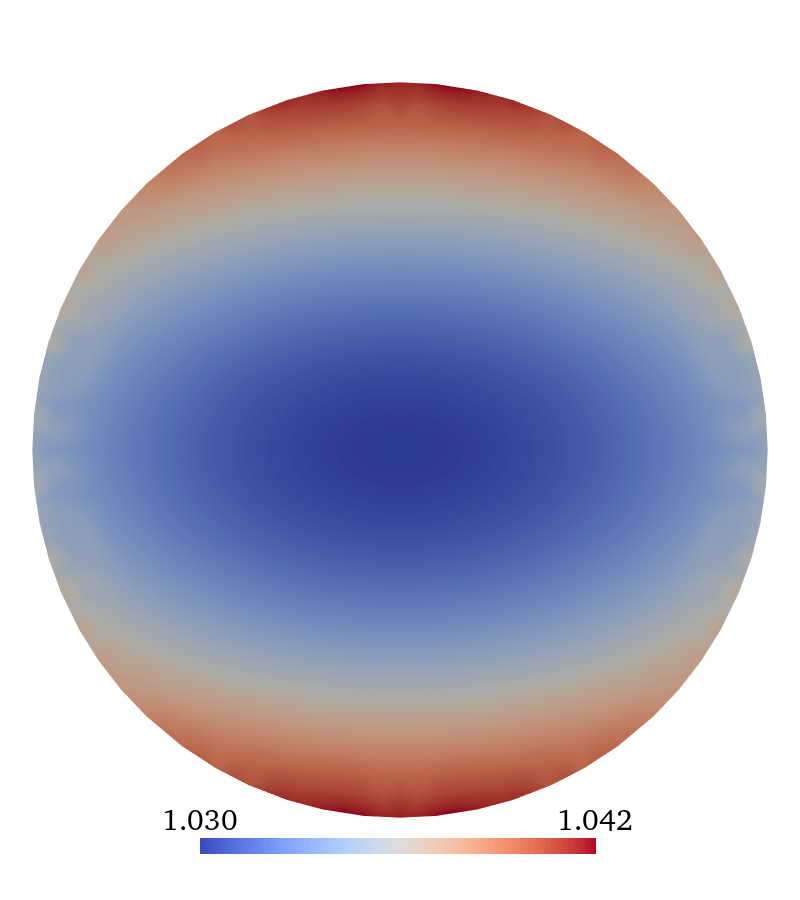}
    }
    \subfloat[$\bar{\mathrm{C}}_{22}$]{
        \includegraphics[width=0.3\textwidth]{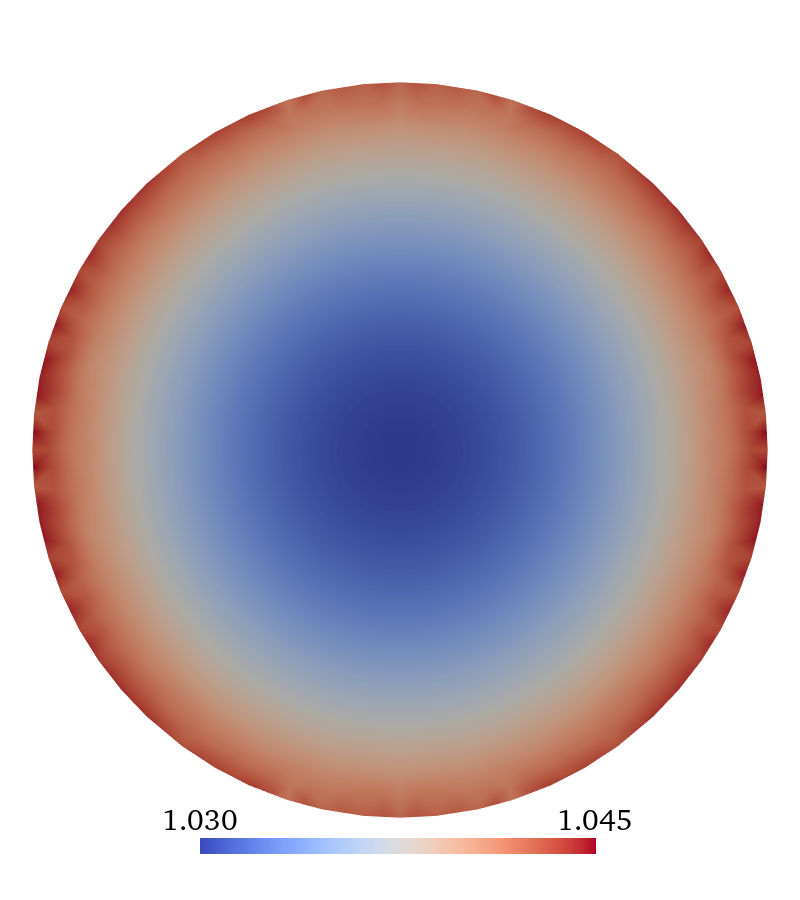}
    }
    \subfloat[$\bar{\mathrm{C}}_{33}$]{
        \includegraphics[width=0.3\textwidth]{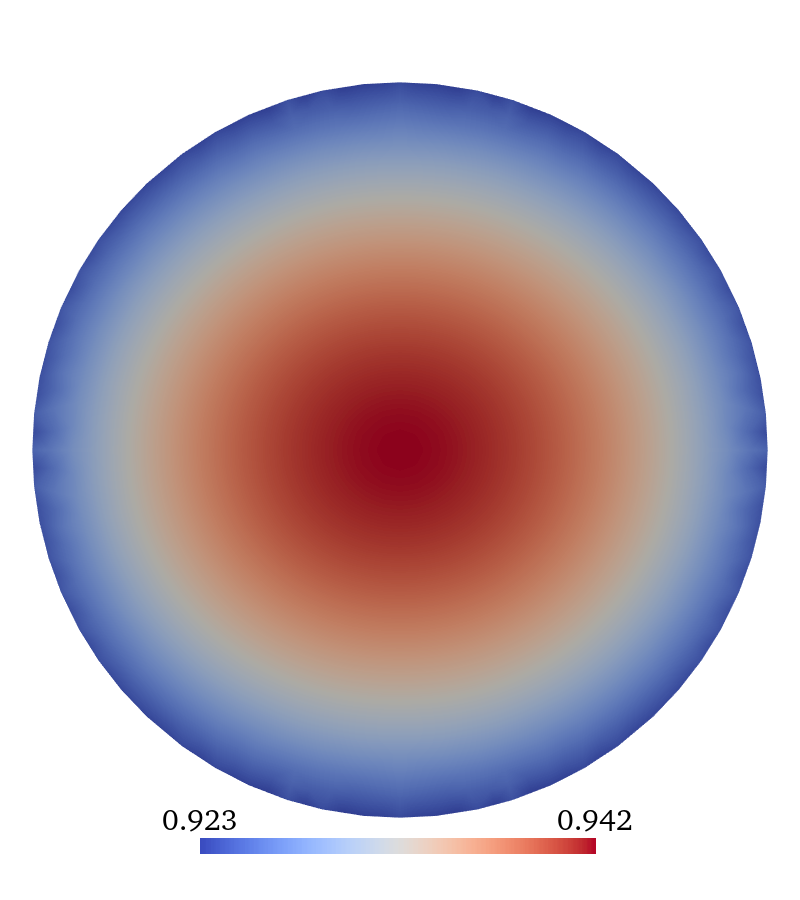}
    }
    \\
    \subfloat[$\bar{\upsigma}_{23} \ / \ \si{\kilo\pascal}$]{
        \includegraphics[width=0.3\textwidth]{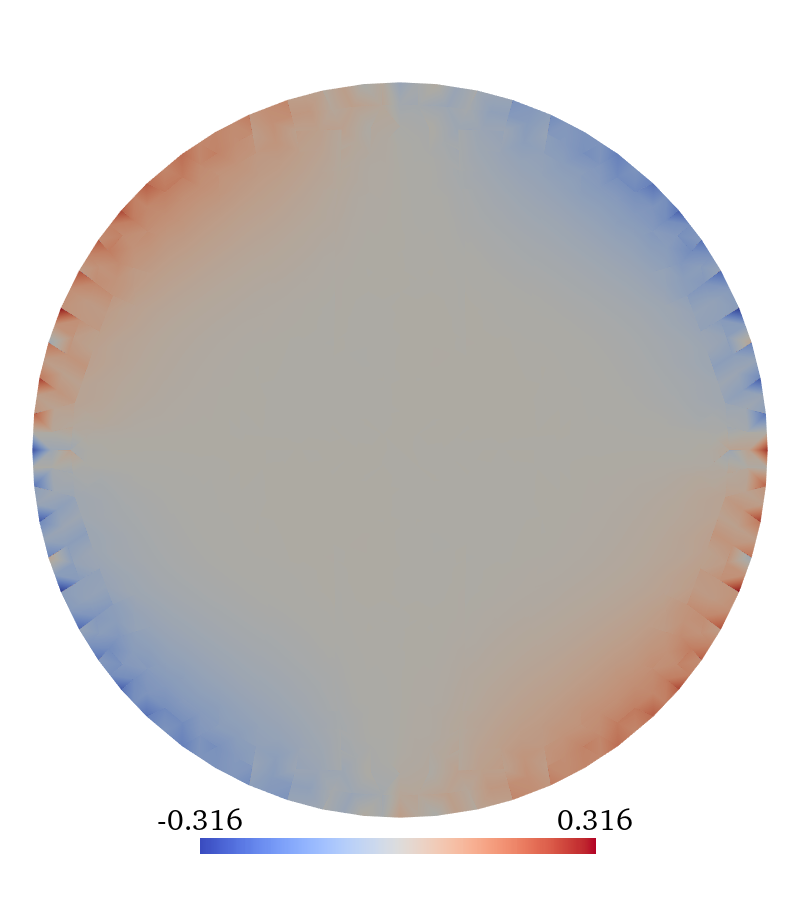}
    }
    \subfloat[$\bar{\mathrm{C}}_{23} \ / \ \si{\kilo\pascal}$]{
        \includegraphics[width=0.3\textwidth]{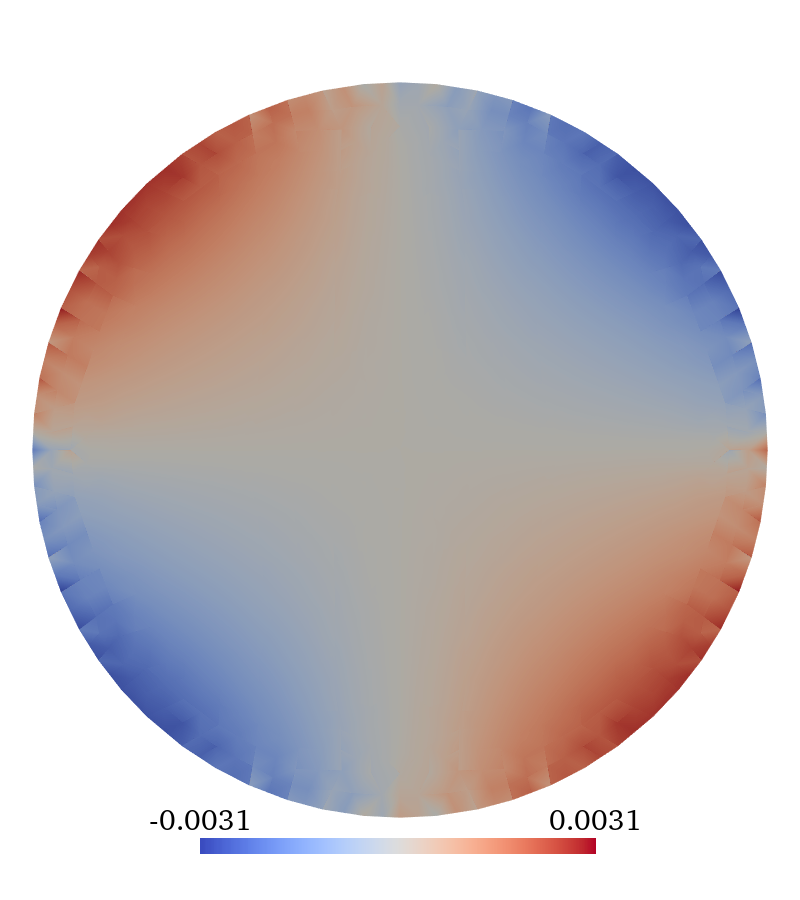}
    }
    \subfloat[$\bar{J}$]{
        \includegraphics[width=0.3\textwidth]{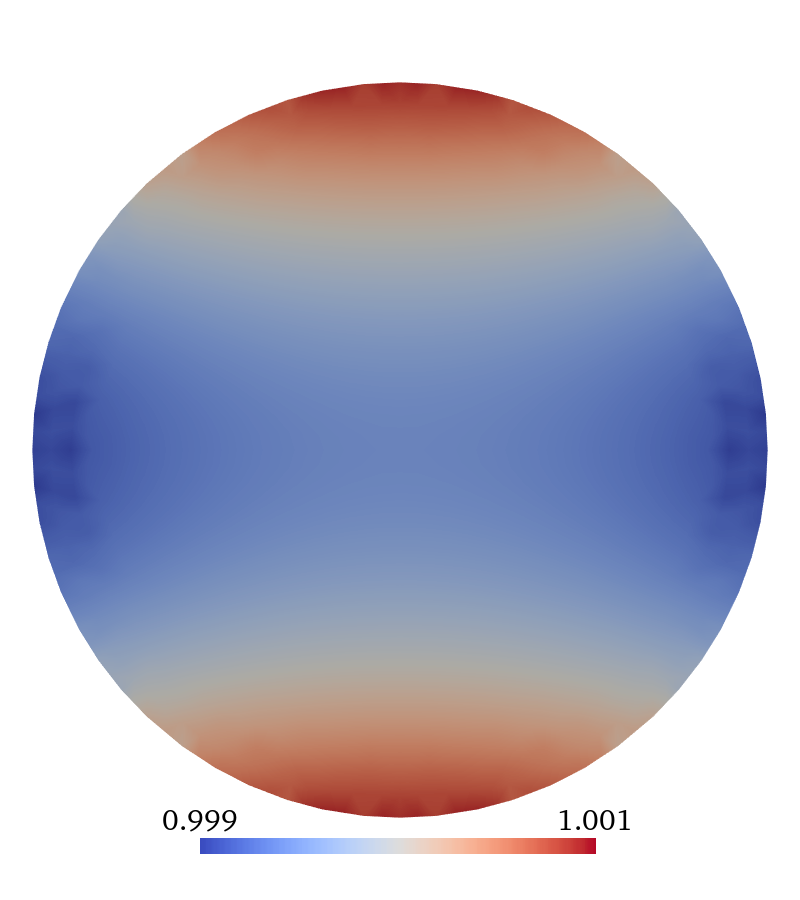}
    }
    \caption{Macroscale fields in the $y$-$z$-plane of a spherical MRE sample with the preferred direction $\bar{\ve{S}} = \ve{e}_3$ for magnetostriction with $\bar{\ve{B}} = (1 \ve{e}_3) \, \si{\tesla}$, as described in Section~\ref{sec:Macroscale_magnetostrictive_effect}.}
    \label{fig:macroscale_fields_magnetostriction}
\end{figure*}

Figure~\ref{fig:macroscale_fields_magnetostriction} presents selected macroscale fields for the magnetostriction load case described in Section~\ref{sec:Macroscale_magnetostrictive_effect}.
The qualitative behavior of the mechanical stress $\bar{\te{\upsigma}}$ and the right Cauchy–Green deformation tensor $\bar{\te{C}}$ resembles that observed in the 2D simulations of a quasi-incompressible isotropic MRE by Kalina \cite{Kalina21Diss}.
Whereas Kalina reports both extension and contraction in $\bar{\mathrm{C}}_{22}$ and $\bar{\mathrm{C}}_{33}$, respectively, we observe only extension in $\bar{\mathrm{C}}_{22}$ and contraction in $\bar{\mathrm{C}}_{33}$.
Compared to the model of compressible isotropic MREs by Gebhart \cite{Gebhart24Diss}, the components of $\bar{\te{C}}$ exhibit an approximately concentric distribution, while the distribution of $\bar{\te{\upsigma}}$ remains qualitatively similar.

\clearpage
\bibliographystyle{unsrtnat} 
\bibliography{PANN_aniso.bib}

\end{document}